\documentclass[pdflatex,sn-basic]{sn-jnl}

\usepackage{threeparttable}
\usepackage{array}
\usepackage{graphicx}%
\usepackage{multirow}%
\usepackage{amsmath,amssymb,amsfonts}%
\usepackage{amsthm}%
\usepackage{mathrsfs}%
\usepackage[title]{appendix}%
\usepackage{xcolor}%
\usepackage{textcomp}%
\usepackage{manyfoot}%
\usepackage{booktabs}%
\usepackage{algorithm}%
\usepackage{algorithmicx}%
\usepackage{algpseudocode}%
\usepackage{listings}%
\usepackage{booktabs}
\usepackage{tabularx}
\usepackage{fancyhdr}
\usepackage{hyperref}
\usepackage{float}
\usepackage{caption}
\usepackage{placeins}
\usepackage{tikz}
\usetikzlibrary{
   shapes.geometric, arrows.meta, positioning, fit, backgrounds, calc, decorations.pathreplacing}

\definecolor{cbluebg}{RGB}{240, 246, 255}
\definecolor{cblueborder}{RGB}{147, 182, 230}
\definecolor{cbluehead}{RGB}{24, 90, 166}

\definecolor{cgreenbg}{RGB}{241, 248, 243}
\definecolor{cgreenborder}{RGB}{152, 203, 168}
\definecolor{cgreenhead}{RGB}{34, 114, 53}

\definecolor{corangebg}{RGB}{255, 247, 240}
\definecolor{corangeborder}{RGB}{245, 190, 150}
\definecolor{corangehead}{RGB}{184, 73, 20}

\definecolor{cpurplebg}{RGB}{250, 245, 255}
\definecolor{cpurpleborder}{RGB}{205, 175, 230}
\definecolor{cpurplehead}{RGB}{112, 53, 153}

\definecolor{cfact}{RGB}{210, 232, 255}
\definecolor{cratio}{RGB}{215, 238, 222}
\definecolor{cprec}{RGB}{255, 228, 200}
\definecolor{carg}{RGB}{235, 215, 250}
\definecolor{cstat}{RGB}{255, 210, 215}
\definecolor{cpres}{RGB}{205, 235, 238}

\theoremstyle{thmstyleone}%
\theoremstyle{thmstyletwo}%

\theoremstyle{thmstylethree}%
\begin{document}

\title{A Study of Prior Case Retrieval Using Lexical, Semantic, and
Rhetorical Role Information in Indian Legal Documents}


\author*[1]{\fnm{Sayed Ayaan} \sur{Ahmed Sha}}\email{sayedayaan810@gmail.com}

\author[1]{\fnm{Sangeetha} \sur{Sivanesan}}

\author[2]{\fnm{Anand Kumar} \sur{Madasamy}}

\author[2]{\fnm{Navya} \sur{Binu}}

\author[1]{\fnm{Aniket} \sur{Mani}}
\author[1]{\fnm{Rishu} \sur{Kumar}}

\author[3]{\fnm{Vasantha} \sur{Kumar}}

\affil[1]{\orgname{National Institute of Technology}, \city{Tiruchirappalli}, \state{Tamil Nadu}, \country{India}}

\affil[2]{\orgname{National Institute of Karnataka}, \city{Surathkal}, \state{Karnataka}, \country{India}}
\affil[3]{\orgname{ Dr. Ambedkar Law University}, \city{Chennai}, \state{Tamil Nadu}, \country{India}}


\abstract{For retrieving prior cases in Indian legal judgments, the problem involves distinguishing relevant legal facts from mere lexical similarities because a prior case that shares a statute with the query judgment is not necessarily relevant. In this paper, we describe an empirical evaluation of three consecutive designs of retrieval systems for the IL-PCR(Indian Legal Prior Case Retrieval) task. We demonstrate that the combination of the rhetorical roles (Fact, Ratio Of The Decision, Precedent, Argument, Statute) is better than either using individual roles or performing full text retrieval, that statute similarity is not discriminative, and that a legal entailment reranker with training data produced by an LLM is much worse in terms of official evaluation than its internal validation score. Two rankers, with excellent internal MRR up to 0.97, performed poorly in terms of official evaluation (MRR as low as 0.14). Thus, we propose a new design with query disjoint splitting and frozen validation fusion. The result is a four stage pipeline with Micro F1 0.2549, MRR 0.6081, and nDCG@10 0.4187.
}




\maketitle

\section{Introduction}\label{sec:introduction}
Identifying earlier judgments that may support the decision in a current case is an important part of legal practice. However, finding such judgments manually from a large collection is time-consuming and often depends on the lawyer's knowledge of previous cases. Prior case retrieval is therefore an information retrieval task that cannot be effectively handled using keyword matching alone.

This task involves several challenges. First, Indian judgments are often lengthy and repetitive, and a single judgment may contain different types of information, such as case facts, arguments, statutes, references to earlier cases, legal reasoning, and the final decision. A prior judgment that refers to the same statute as the current case may not be legally relevant. Similarly, two judgments may discuss the same legal issue but use very different words to express it.

This research studies how different retrieval and reranking methods can be combined to improve prior case retrieval for Indian Supreme Court judgments. It examines how lexical, semantic, and rhetorical-role information can help identify and rank relevant prior cases.

\section{Related Work}\label{sec:related work}
Initial information retrieval systems for legal information used the technique of lexical matching \citep{luhn1957statistical, cooper1971definition, salton1988term}. Previous investigations into the use of prior cases through information retrieval also made use of the feature based method, involving the application of Support Vector Machines (SVMs) \citep{al2001machine, jackson2003information}. \citep{robertson2009probabilistic} proposed the Probabilistic Relevance Framework (PRF), which models the estimation of relevance of documents given a query. Through this framework, BM25 emerged as an effective ranking function and further developments were made through BM25F, taking into account the structure of the document and multiple fields. The FIRE 2019 AILA track explored methods for retrieving relevant prior cases \citep{bhattacharya2019overview}. Similarly, COLIEE 2021 showed that BM25 provides a strong baseline for legal case retrieval. Participants combined BM25 and XGBoost with techniques such as TF-IDF, large language models, and BERT-based transformers \citep{rabelo2022overview}. TLIR at COLIEE 2021 combined lexical retrieval with neural reranking in a two-stage framework \citep{ma2021retrieving,rabelo2022overview}. BM25 and TF-IDF retrieved candidate cases, which were then reranked using BERT-PLI to capture semantic relationships between query and candidate cases. Another approach extracted evidential sentences representing key facts and legal reasoning from judgments \citep{ali2021prior}. Instead of comparing entire documents, it matched evidence from query cases with evidence from prior cases. The results showed that this evidence-based approach outperformed traditional full-text similarity methods in retrieving relevant precedents \citep{ali2021prior}. \textbf{LexGLUE} provides a benchmark of seven English-language datasets for evaluating natural language understanding models across legal tasks. For prior case retrieval, \citep{joshi-etal-2023-u} proposed an event-based approach that extracts events from query and candidate documents. Each event consists of a predicate describing an action and its arguments representing the participants. Using spaCy's dependency parser, the approach identifies these events, computes their similarities through an interaction matrix, and ranks candidate cases accordingly.

\citep{joshi-etal-2024-il} introduced \textbf{IL-TUR}, a benchmark covering eight Indian legal NLP tasks in English and nine Indian languages. It includes tasks such as case retrieval, translation, and information extraction using full-length court judgments. \citep{nguyen2024attentive} proposed two neural models for legal case retrieval: Attentive CNN, which uses convolutional networks, and Paraformer, which uses a pretrained Transformer. Both models encode sentences and combine their representations to assess the relevance of candidate legal documents to a query. In COLIEE 2024, researchers explored a multistage retrieval framework combining lexical matching, semantic similarity, and ranking techniques for legal case retrieval \citep{goebel2024overview,li2024towards}. The framework also used preprocessing to reduce noise and retain legally relevant information. However, evaluating retrieval models on full court judgments can create an unrealistic setting for assessing their practical performance. \citep{nigam2025segment} introduced TraceRetrieve, a pipeline that reflects real-world legal searches by using key sections, such as Facts and Issues, instead of entire judgments. Queries combining Facts, Issues, and Reasoning outperformed full-document queries, achieving a higher MAP (0.3783 vs. 0.3484). The pipeline combines lexical retrievers (BM25 \citep{robertson2009probabilistic}, QLD, and BM25-ngram) with semantic models (SAILER \citep{li2023sailer} and DELTA \citep{li2025delta}). LightGBM then combines their scores to rank candidate cases, followed by date filtering and dynamic cut-offs to refine the results. CaseGNN addresses the challenges of representing long legal documents while preserving their structure. It uses a Text-Attributed Case Graph (TACG) and an Edge Graph Attention Layer (EdgeGAT) to learn case representations \citep{tang2024casegnn}. It ranks cases by comparing their embeddings, with contrastive learning and hard negatives helping distinguish relevant precedents from irrelevant ones \citep{tang2024casegnn}. Results on the IL-PCR benchmark \citep{joshi-etal-2024-il} suggest that Transformer and LLM based methods still face challenges in Indian prior case retrieval, motivating further exploration of NLP-based approaches.

\section{Proposed Methodology}\label{sec:methodology}

This study is organised into three experimental phases and a four-stage retrieval pipeline. The experimental phases represent the progression of the research: Phase 1 (Baseline Study) examines lexical and rhetorical-role-based retrieval; Phase 2 (Entailment Study) investigates semantic relationships between query facts and candidate legal reasoning; and Phase 3 (Transfer Study) explores role-based neural ranking and the reliability of internal model evaluation. These phases are described in Sections~\ref{sec:phase 1}, \ref{entailment}, and \ref{sec:role analysis}, respectively.

The proposed pipeline brings together the methodological developments from these experiments. Phase 1 informs Stage 1 (Role-Aware Candidate Retrieval), Phase 2 informs Stage 2 (Facts-to-Ratio Semantic Reranking), and Phase 3 informs Stage 3 (Directed Role-Pair Ranking) as well as the validation-based model selection used in Stage 4 (Final Rank Fusion). Thus, the phases describe the experimental progression, while the stages represent the components of the final retrieval system.

The pipeline is motivated by the fact that lexical similarity alone may not adequately capture legal relevance. Two judgments may discuss similar legal principles using different expressions, while sharing terms does not necessarily mean that they address the same legal issue. Furthermore, different sections of a judgment serve different purposes. We therefore combine lexical matching, semantic comparison, and rhetorical-role information to capture different aspects of legal relevance.

The first phase uses sparse lexical retrieval through eleven BM25 channels: one based on the full text and ten based on selected query--candidate rhetorical-role pairs. Although TF-IDF performed better than BM25 in the official baseline evaluation, BM25 was retained for the multi-channel architecture to maintain a common retrieval framework across the full-text and role-specific channels. The channel rankings are combined using Reciprocal Rank Fusion (RRF), with uniform weights and (K=60). Candidate diversification is then applied to construct a pool of 600 judgments. This design follows the finding that combining multiple role-specific retrieval channels provides a stronger retrieval signal than relying on a single role pair.

The second phase compares the query's Facts section with the candidate judgment's Ratio of the Decision section. InLegalBERT is used as a cross-encoder to assess their semantic relationship. To handle long judgments, query Facts are represented using a 192-token head--tail selection, while candidate Ratio passages are divided into overlapping windows with a 128-token overlap. Window-level scores are combined using log-mean-exp aggregation. The resulting semantic ranking is fused with the lexical ranking from pipeline Phase 1 using RRF, with a neural weight of 2.0. The top 200 candidates are retained for the next phase. This design follows the finding that semantic reranking alone does not reliably generalise to official relevance labels, whereas combining semantic and lexical evidence can improve retrieval performance.

The third phase extends the comparison to ten selected, directional rhetorical-role pairs because legal relevance may involve relationships beyond Facts-to-Ratio. For each query–candidate pair, InLegalBERT generates ten role-pair representations. A shared projection layer reduces their dimensionality from 768 to 128, while role-identity vectors preserve information about the type of relationship. The model estimates a signed contribution for each pair, allowing it to represent both positive and negative relevance evidence. Masked gates control these contributions when role information is unavailable. The phase 2 semantic score and role-coverage information are also included. A ranking multilayer perceptron (MLP) combines these features to produce a single relevance score for each candidate. The candidates are then sorted according to these scores to obtain the phase 3 ranking.

The final phase combines the rankings from phases 2 and 3 using RRF. These rankings provide complementary evidence: Phase 2 captures the semantic relationship between query facts and candidate legal reasoning, while Phase 3 considers relationships across multiple rhetorical roles. Since the models use different scoring mechanisms, rank-based fusion avoids directly comparing their raw scores. Equal weights are used with \(K=30\), and the final top-9 cutoff is selected using validation data. The fusion configuration is then frozen before official evaluation to prevent test-set information from influencing model selection. The resulting ranking provides the final shortlist of prior judgments for each query.

Figure \ref{fig:architecture} provides a detailed overview of the workflow adopted

\begin{figure}[!tbp]
\centering
\resizebox{0.85\textwidth}{!}{%
\begin{tikzpicture}[
    font=\sffamily\footnotesize,
    >={Stealth[length=2mm]},
    stagebox/.style={draw, rounded corners=3pt, line width=0.7pt, align=center,
                     inner sep=3pt, minimum width=3.7cm},
    sub/.style={draw, rounded corners=2pt, line width=0.4pt, fill=white, align=center,
                inner sep=2pt, text width=3.3cm, minimum height=0.65cm},
    io/.style={draw, rounded corners=3pt, line width=0.7pt, align=center,
               inner sep=3pt, text width=2.5cm},
    arrow/.style={->, line width=0.7pt},
    stagetitle/.style={font=\sffamily\footnotesize\bfseries, align=center, text width=3.5cm}
]

\node[io, fill=blue!6, minimum height=0.9cm] (q) at (0,-1.6) {Query judgment $q$};
\node[io, fill=blue!6, below=0.4cm of q] (roles)
  {Rhetorical-role representation\\[1pt]
   {\scriptsize Fact, Ratio, Precedent, Argument, Statute, Present-Court Ruling}};

\node[stagetitle] (t1) at (4.5,0) {\textbf{1}~Role-aware\\candidate retrieval};
\node[sub, fill=blue!5, below=0.15cm of t1] (s1a) {Full-text BM25 channel};
\node[sub, fill=blue!5, below=0.15cm of s1a] (s1b)
  {10 directed role-pair BM25 channels\\{\scriptsize (Fact--Fact, Fact--Ratio, $\dots$, Statute--Ratio)}};
\node[sub, fill=blue!5, below=0.15cm of s1b] (s1c)
  {RRF over 11 channels\\{\scriptsize uniform weights, $K{=}60$}};
\node[sub, fill=blue!15, below=0.15cm of s1c] (s1d) {Diversified \textbf{top-600} pool};
\begin{scope}[on background layer]
  \node[stagebox, draw=blue!60!black, fill=blue!3, fit=(t1)(s1a)(s1b)(s1c)(s1d)] (b1) {};
\end{scope}

\node[stagetitle] (t2) at (9.0,0) {\textbf{2}~Sliding-window\\entailment reranking};
\node[sub, fill=green!6, below=0.15cm of t2] (s2a)
  {Query Facts {\scriptsize (192-token head--tail)} $+$ candidate Ratio windows {\scriptsize (128-token overlap)}};
\node[sub, fill=green!6, below=0.15cm of s2a] (s2b)
  {InLegalBERT cross-encoder\\{\scriptsize log-mean-exp pooling (Eq.~9)}};
\node[sub, fill=green!6, below=0.15cm of s2b]  (s2d)
  {RRF: retrieval rank $+$ neural rank\\{\scriptsize neural weight 2.0}};
\node[sub, fill=green!15, below=0.15cm of s2d] (s2e) {\textbf{Top-200}\\{\scriptsize entailment (Stage~2) rank}};
\begin{scope}[on background layer]
  \node[stagebox, draw=green!50!black, fill=green!3, fit=(t2)(s2a)(s2b)(s2d)(s2e)] (b2) {};
\end{scope}

\node[stagetitle] (t3) at (9.0,-7.4) {\textbf{3}~Directed role-pair\\ranker};
\node[sub, fill=orange!8, below=0.15cm of t3] (s3a)
  {10 role-pair embeddings\\{\scriptsize cached, fine-tuned InLegalBERT}};
\node[sub, fill=orange!8, below=0.15cm of s3a] (s3b)
  {Projection 768$\to$128 $+$ role identity vector};
\node[sub, fill=orange!8, below=0.15cm of s3b] (s3c)
  {Signed contribution in $[-1,1]$ $+$ masked gate};
\node[sub, fill=orange!8, below=0.15cm of s3c] (s3d)
  {Ranking MLP\\{\scriptsize $+$ Stage~2 score, role coverage}};
\node[sub, fill=orange!15, below=0.15cm of s3d] (s3e) {Role-aware rank};
\begin{scope}[on background layer]
  \node[stagebox, draw=orange!70!black, fill=orange!3, fit=(t3)(s3a)(s3b)(s3c)(s3d)(s3e)] (b3) {};
\end{scope}

\node[stagetitle] (t4) at (4.5,-7.4) {\textbf{4}~Validation-frozen\\fusion (RRF)};
\node[sub, fill=violet!7, below=0.15cm of t4] (s4a) {Entailment (Stage~2) rank\\{\scriptsize $w_b{=}1$}};
\node[sub, fill=violet!7, below=0.15cm of s4a] (s4b) {Role-aware rank\\{\scriptsize $w_r{=}1$}};
\node[sub, fill=violet!7, below=0.15cm of s4b] (s4c) {RRF, $K{=}30$\\{\scriptsize selected on validation}};
\node[sub, fill=violet!15, below=0.15cm of s4c] (s4d) {Frozen cutoff: \textbf{top-9}};
\begin{scope}[on background layer]
  \node[stagebox, draw=violet!70!black, fill=violet!3, fit=(t4)(s4a)(s4b)(s4c)(s4d)] (b4) {};
\end{scope}

\node[io, fill=gray!10, minimum height=1.4cm] (out) at (0,-11.4)
  {\textbf{Output}\\Top-9 prior cases per query};

\draw[arrow] (q.east)     -- ++(0.5,0) |- (b1.west);
\draw[arrow] (roles.east) -- ++(0.5,0) |- (b1.west);
\draw[arrow] (b1.east |- b2.center) -- (b2.west |- b2.center);
\draw[arrow] (b2.south) -- (b3.north);
\draw[arrow] (b3.west |- b4.center) -- (b4.east |- b4.center);
\draw[arrow] (b4.west |- out.center) -- (out.east);
\draw[arrow, dashed] ([xshift=0.4cm]b2.south west) -- ([xshift=-0.4cm]b4.north east);

\end{tikzpicture}%
}
\caption{Overview of the four-stage pipeline: (1) eleven-channel BM25 candidate retrieval, (2) sliding-window InLegalBERT entailment reranking, (3) directed role-pair ranker, and (4) validation-frozen RRF fusion. The dashed arrow shows that the Stage~2 rank also feeds the final fusion.}
\label{fig:architecture}
\end{figure}

\section{Dataset and Evaluation Metrics}\label{secc:dataset and metrics}

This study uses the IL-PCR (Indian Legal Prior Case Retrieval) collection \citep{joshi-etal-2024-il}, which contains 7,070 English-language legal documents. The collection provides separate query and candidate judgments, along with relevance judgments that identify the prior cases cited in each query. The training set contains 827 queries and 4,320 candidate judgments, while the official test set contains 237 queries and 1,727 candidate judgments. The dataset statistics are presented in Table \ref{tab:query_candidate_counts}

\begin{table}[htbp]
  \centering
  \caption{Query vs Candidate counts per split}
  \label{tab:query_candidate_counts}
  \begin{tabular}{lrr}
    \hline
    \textbf{Split} & \textbf{Query} & \textbf{Candidate} \\
    \hline
    dev   & 118 & 1,023 \\
    test  & 237 & 1,727 \\
    train & 827 & 4,320 \\
    \hline
    \textbf{Total} & \textbf{1,182} & \textbf{7,070} \\
    \hline
  \end{tabular}
\end{table}

\section{Phase 1: Baseline Study}\label{sec:phase 1}
\subsection{Full-text lexical baselines}\label{subsec:lexical}

Each query was cleaned, indexed, ranked against the candidate collection, and evaluated using a common routine that accounted for document identifiers. The experiments compared full-text BM25 and TF-IDF, TextRank summaries \citep{mihalcea2004textrank}, role-aware sparse retrieval, MiniLM role embeddings \citep{wang2020minilm}, InLegalBERT dense retrieval \citep{paul2023pre}, and lexical--semantic hybrid methods. TextRank is an unsupervised, graph-based algorithm that ranks sentences based on their similarity to other sentences using a PageRank-style approach to generate shorter case summaries.

We used BM25 \citep{robertson2009probabilistic} as the baseline for prior case retrieval because exact term matches are important in legal texts, particularly for citations, legal terminology, and statutory language. Judgments were tokenized by splitting the text on whitespace and indexed using BM25Okapi. For each query judgment, BM25 scores were calculated against all candidate judgments, and the top-$k$ candidates were retained for evaluation. 




We also used TF-IDF with cosine similarity as another lexical baseline. It included unigram features, removed English stopwords, applied sublinear term frequency scaling, and used a maximum vocabulary size of 100,000 features. Candidate judgments and query judgments were represented using the same fitted TF-IDF vectorizer, and cosine similarity was used to rank the candidates. 

\begin{figure}[t]
    \centering
    \includegraphics[width=1.1\textwidth]{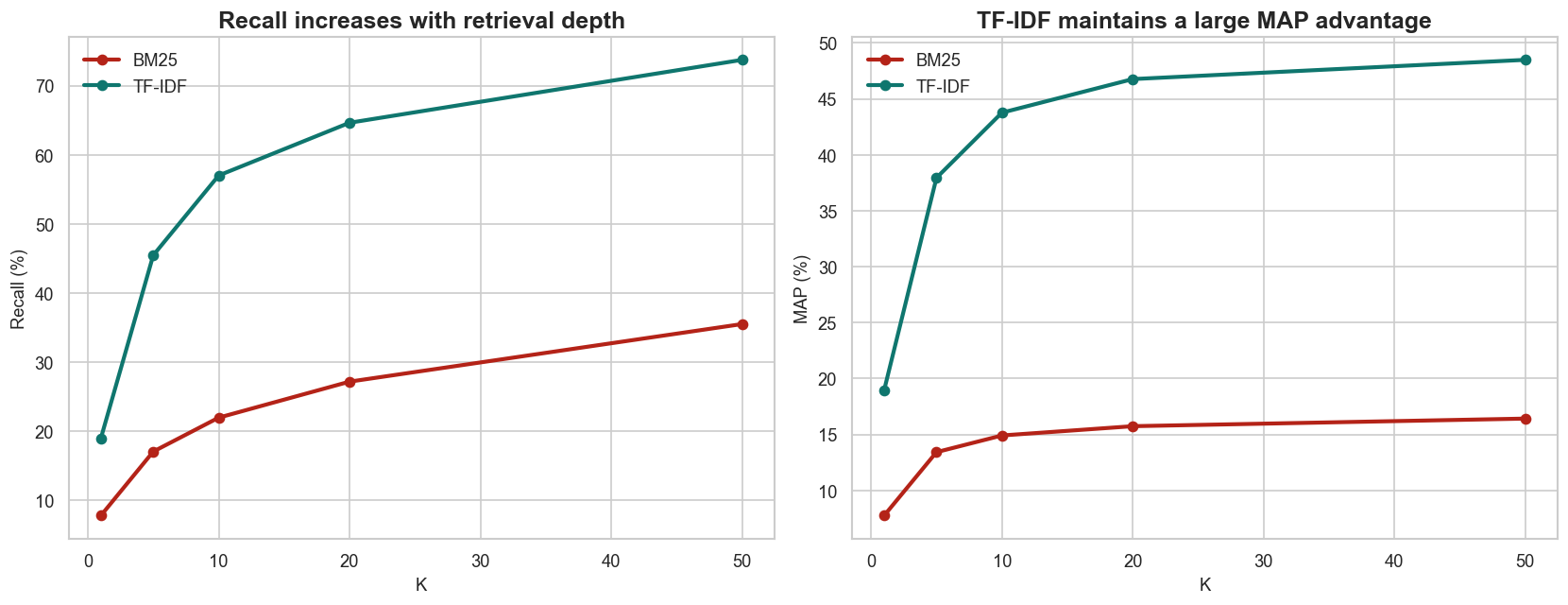}
    \caption{Recall and MAP of the reproduced full text BM25 and TF-IDF baselines}
    \label{fig:bm25_tfidf}
\end{figure}

Table~\ref{tab:fulltext-k10} shows the results at $K=10$. TF-IDF outperformed BM25 across all five metrics, making it the stronger full-text lexical baseline in our experiments. Both methods used the same candidate identifiers, query relevance judgments, and retrieval depths. Self-matches were also removed from both methods because a query judgment should not retrieve itself as a prior case. Figure~\ref{fig:bm25_tfidf} shows that TF-IDF maintained its advantage as retrieval depth increased. At $K=50$, TF-IDF achieved a Recall of 0.7375 and MAP of 0.4848, compared with 0.3549 and 0.1642, respectively, for BM25.

\begin{table}[t]
\centering
\small
\setlength{\tabcolsep}{10.5pt}
\begin{tabular}{lccccc}
\toprule
Method & Precision & Recall & MRR & MAP & nDCG \\
\midrule
BM25   & 0.1239 & 0.2194 & 0.4252 & 0.1492 & 0.2444 \\
TF-IDF & \textbf{0.3154} & \textbf{0.5700} & \textbf{0.7385}
       & \textbf{0.4376} & \textbf{0.5932} \\
\bottomrule
\end{tabular}
\caption{Full text retrieval results at $k=10$, computed on the 237 query official test collection against all 1,727 candidates}
\label{tab:fulltext-k10}
\end{table}

\subsection{Rhetorical Role Coverage and Single Role Retrieval}\label{subsec:role coverage}

Legal decisions consist of various forms of information that include the facts of the case, arguments put forward by the parties, statutes applicable, precedents cited and the reason for the decision of the court. Each of the forms serves a different role in the decision-making process. The rhetorical roles assist in recognizing and organizing the various parts based on the roles played in a legal document. Some examples are that the Facts part explains the details of the case, the Argument part provides the arguments of the parties while the Ratio of the Decision part gives the reasons for the decision of the court.

Rhetorical roles are useful for Prior Case Retrieval (PCR) because finding a relevant earlier judgment involves more than matching words between two documents. A candidate judgment may be relevant because its legal reasoning applies to the facts of the query case, even when both judgments use different words. Similarly, two judgments may contain similar words but deal with different legal issues. Comparing specific rhetorical roles helps identify these connections more clearly. For example, comparing the Facts of a query judgment with the Ratio of the Decision of a candidate judgment can help find earlier cases whose legal reasoning is relevant to the current case. Comparing the Precedent sections can also help identify judgments that refer to similar earlier decisions.

\citep{malik2022semantic} introduced a corpus of Indian legal documents annotated with 13 rhetorical roles, building on the seven-role scheme proposed by \citep{bhattacharya2019identification}: Facts (FAC), Ruling by Lower Court (RLC), Argument (ARG), Statute (STA), Precedent (PRE), Ratio of the Decision (Ratio), and Ruling by Present Court (RPC). These annotations make it possible to compare specific parts of judgments instead of relying only on their full texts. Tables~\ref{query_role_distribution} and~\ref{candidate_role_distribution} show the distribution of rhetorical roles in the query and candidate judgments.

We first examined whether comparing individual role pairs could improve PCR. We used the rhetorical-role annotations already available in the IL-PCR dataset and did not apply an additional classifier. Some judgments did not contain text for every role. Therefore, we excluded a role-pair comparison whenever either judgment lacked the required role. We evaluated five role pairs. Precedent–Precedent achieved the best MRR, MAP, and nDCG@10, while Fact–Fact achieved slightly higher Recall@10. Issue–Ratio of the Decision showed poor performance. The results are presented in Table~\ref{tab:role-pairs} .

Next, we examined whether combining multiple role-pair similarities could help identify relevant prior cases. We created query–candidate pairs from the IL-PCR dataset and calculated similarity features between corresponding rhetorical roles in each pair. This produced 10,362 pairs: 1,727 relevant pairs and 8,635 non-relevant pairs, based on the IL-PCR annotations. We divided these pairs into training and validation sets by query, keeping all pairs belonging to the same query in one set. This prevented the same query from appearing in both sets. We then trained a logistic regression model using the role-pair similarity features. The model achieved an ROC-AUC of 0.7546 on the validation set. We also examined the learned coefficients to understand which role pairs contributed most to the model. Fact–Fact had the highest coefficient, followed by Ratio–Ratio and Precedent–Precedent. This finding differed from the individual role-pair experiment, where Precedent–Precedent performed best. When multiple role-pair features were considered together, Fact–Fact and Ratio–Ratio received greater importance.

\begin{figure}[t]
    \centering
    \includegraphics[width=1.1\textwidth]{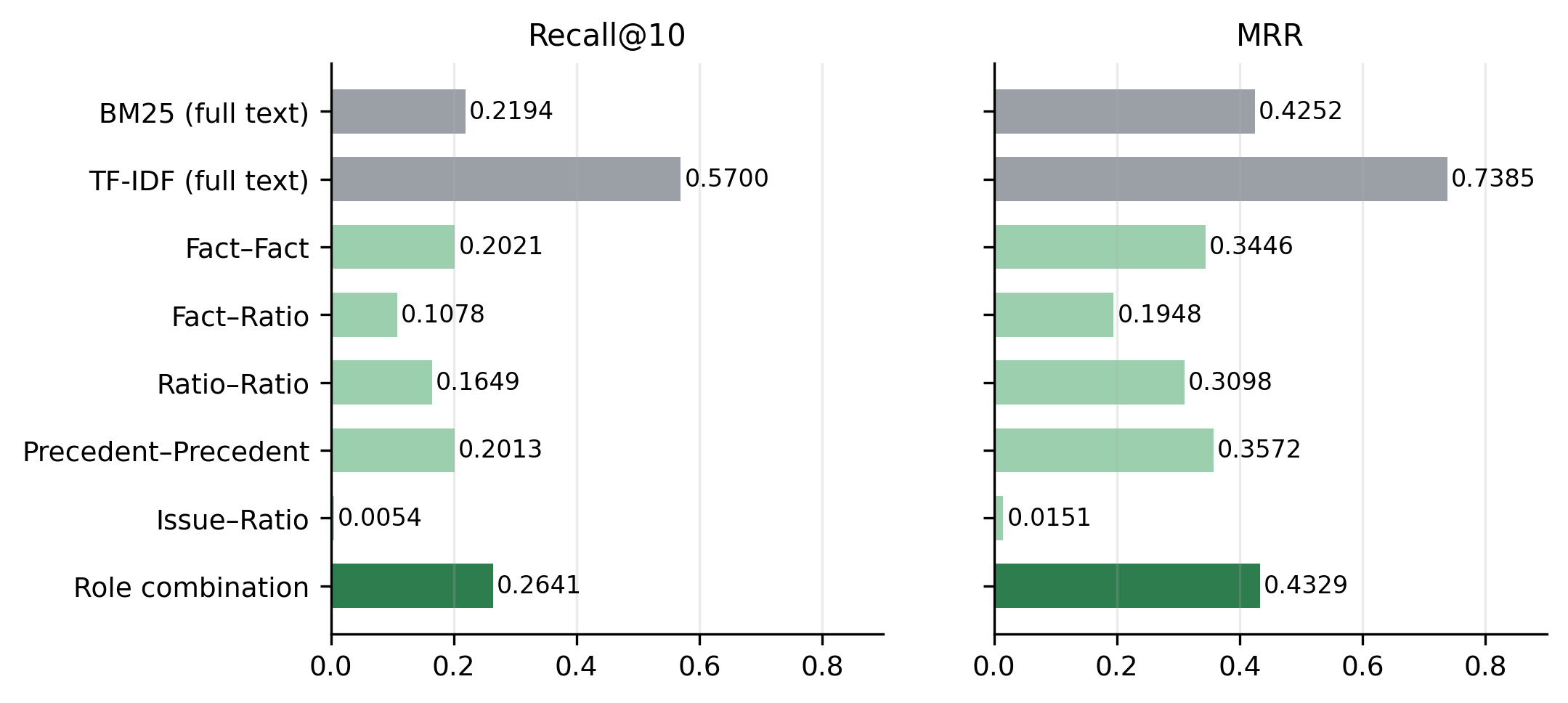}
    \caption{Comparison of full-text and rhetorical-role-based retrieval configurations using Recall@10 and Mean Reciprocal Rank (MRR).}
    \label{fig:phase1_roles}
\end{figure}

\begin{table}[ht]
\centering
\caption{Query Rhetorical Role Distribution}
\label{query_role_distribution}
\begin{tabular}{lrr}
\hline
\textbf{Rhetorical Role} & \textbf{Count} & \textbf{Percentage} \\
\hline
Precedent              & 21025 & 38.35\% \\
RatioOfTheDecision     & 14247 & 25.98\% \\
Fact                   & 12785 & 23.32\% \\
Argument               & 3163  & 05.77\%  \\
Statute                & 2467  & 04.49\%  \\
RulingByPresentCourt   & 652   & 01.19\% \\
RulingByLowerCourt     & 498   & 0.90\% \\
\hline
\end{tabular}
\end{table}

\begin{table}[ht]
\centering
\caption{Candidate Rhetorical Role Distribution}
\label{candidate_role_distribution}
\begin{tabular}{lrr}
\hline
\textbf{Rhetorical Role} & \textbf{Count} & \textbf{Percentage} \\
\hline
Precedent              & 168022 & 39.98\% \\
RatioOfTheDecision     & 112820 & 26.85\% \\
Fact                   & 90518 & 21.53\% \\
Argument               & 22729  & 05.41\%  \\
Statute                & 18887  & 04.49\%  \\
RulingByPresentCourt   & 4866  & 01.16\%  \\
RulingByLowerCourt     & 2447   & 0.58\%  \\
\hline
\end{tabular}
\end{table}

\begin{table}[t]
\centering
\scriptsize
\setlength{\tabcolsep}{10.5pt}
\begin{tabular}{lcccccc}
\toprule
Role pair & R@1 & R@5 & R@10 & MRR & MAP & nDCG@10 \\
\midrule
Fact--Fact
    & 0.0600 & 0.1517 & \textbf{0.2021}
    & 0.3446 & 0.1159 & 0.2007 \\
Fact--Ratio
    & 0.0138 & 0.0725 & 0.1078
    & 0.1948 & 0.0504 & 0.1074 \\
Ratio--Ratio
    & 0.0435 & 0.1156 & 0.1649
    & 0.3098 & 0.0941 & 0.1759 \\
Precedent--Precedent
    & \textbf{0.0618} & \textbf{0.1553} & 0.2013
    & \textbf{0.3572} & \textbf{0.1258} & \textbf{0.2092} \\
Issue--Ratio
    & 0.0011 & 0.0034 & 0.0054
    & 0.0151 & 0.0020 & 0.0054 \\
\bottomrule
\end{tabular}
\caption{Individual rhetorical role retrieval results.}
\label{tab:role-pairs}
\end{table}

\section{Legal Entailment Study}\label{entailment}
Phase 2 explores a reranking strategy that assesses how well a candidate case's Ratio of the Decision supports the facts of a query case. We call this relationship Facts-to-Ratio support. It does not represent textual entailment in the traditional natural language inference sense. Instead, it examines whether the legal principles and reasoning in a prior case's Ratio support the circumstances described in the query case. The query provides the facts and conditions underlying the dispute, while the candidate provides the legal reasoning established in a previous case. We hypothesize that candidates whose Ratio better supports the query facts should receive higher reranking scores.

\subsection{Candidate Pool Construction and Role Validation}
Phase 2 used the Indian Legal Prior Case Retrieval (IL-PCR) dataset \citep{joshi-etal-2024-il}. It contains query judgments, candidate judgments, and official relevance labels based on case citations. We used the prepared rhetorical-role records to extract the Facts from each query judgment and the Ratio of the Decision from each candidate judgment.
The Phase 2 experiment used BM25-retrieved candidates rather than the full candidate collection. Data preparation involved five stages.
First, we validated the rhetorical-role text. Of the 827 query records, 799 contained valid Fact text. Of the 4,319 candidate records, 3,882 contained valid Ratio of the Decision text. Second, BM25 retrieved the top 50 candidates for each valid query, producing 39,950 query–candidate pairs. Third, we removed 679 self-matches in which a query judgment was retrieved as its own candidate, leaving 39,271 pairs. Fourth, Gemini labeled each pair to assess whether the candidate's Ratio of the Decision provided legal support for the query's Facts. Finally, we cleaned the labeled data to remove unsuitable records, resulting in 39,074 query–candidate pairs across 795 queries.

\subsection{Direct InLegalBERT Cross-Encoder}
\begin{figure}[t]
    \centering
    \includegraphics[width=1.1\textwidth]{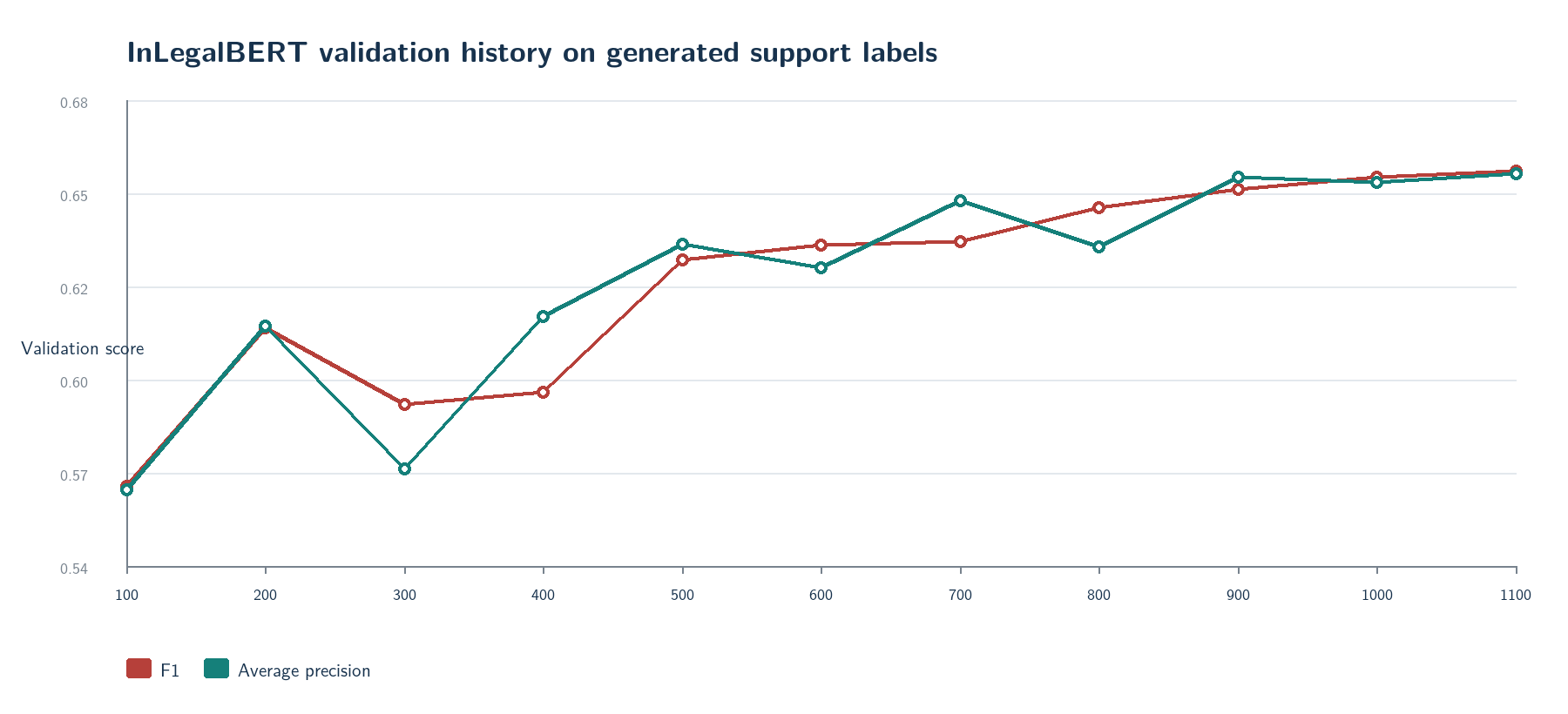}
    \caption{InLegalBERT cross-encoder validation history}
    \label{fig:phase2_initial}
\end{figure}

We first trained an InLegalBERT cross-encoder \citep{paul2023pre} on the Facts-to-Ratio dataset. The experiment used 4,635 query–candidate pairs: 3,708 for training and 927 for validation. Only 14.67\% of the pairs were positive, indicating a class imbalance. Therefore, F1 and Average Precision were more informative evaluation metrics than accuracy alone.
The model received the query's Facts and the candidate's Ratio of the Decision, with a maximum input length of 512 tokens. We trained it for one epoch and evaluated it every 100 steps. As shown in Figure~\ref{fig:phase2_initial}, validation F1 increased from 0.5639 at step 100 to 0.6586 at step 1,100. Average Precision followed a similar trend, rising from 0.5628 to 0.6578.

\begin{table}[t]
\centering
\small
\setlength{\tabcolsep}{8pt}
\renewcommand{\arraystretch}{1.1}
\begin{tabular}{lc}
\toprule
\textbf{Measure} & \textbf{Value} \\
\midrule
BM25 weight       & 0.55 \\
Reranker weight   & 0.45 \\
Correct pairs     & 429 \\
Official gold pairs & 1,727 \\
Precision@10      & 0.181013 \\
Recall@10         & 0.248408 \\
Micro-F1@10       & \textbf{0.209422} \\
\bottomrule
\end{tabular}
\caption{Official Phase 2 evaluation of the BM25--InLegalBERT hybrid at $k=10$.}
\label{tab:phase2-hybrid}
\end{table}

The entailment reranker was first evaluated using the Gemini-generated support labels. It outperformed BM25 on all ranking metrics, achieving an MRR of 0.822895, Recall@10 of 0.973231, MAP of 0.799915, and nDCG of 0.819971. We then tested whether the learned entailment signal transferred to the official case retrieval task using the IL-PCR relevance labels. In contrast to the support-label evaluation, the InLegalBERT reranker alone did not outperform BM25. BM25 achieved a Micro-F1@10 of 0.194777, compared with 0.177 for the reranker. We combined the normalized BM25 and reranker scores to improve retrieval performance. The best configuration used a weight of 0.55 for BM25 and 0.45 for the reranker, achieving a Micro-F1@10 of 0.209422. The detailed results are shown in Table~\ref{tab:phase2-hybrid}.


\subsection{First Hierarchical Sliding Window Branch}
To address the long-input limitation of the direct cross-encoder, we tested a hierarchical approach that split query Facts and candidate Ratios of the Decision into overlapping windows. InLegalBERT encoded the query–candidate window pairs, and an attention mechanism aggregated their representations. The query-disjoint split contained 33,177 training pairs and 5,897 validation pairs. Training was limited to one epoch, yielding a validation MRR of 0.3957. On the official test set of 237 queries, the model achieved P@10 of 0.1705, R@10 of 0.2339, and F1@10 of 0.1972. Although the hierarchical approach handled longer role-specific text, its official F1@10 remained below that of the 55:45 BM25–InLegalBERT hybrid (Micro-F1@10 = 0.209422).

\section{Role Based Ranking}\label{sec:role analysis}
In this phase, we examine how rhetorical-role information can improve prior case retrieval when combined with other ranking methods. We investigate whether certain rhetorical roles are more useful for retrieval and whether the ranking methods studied earlier can make effective use of this information.

\subsection{Positive and Negative Importance}
Phase 3 uses role supervision to distinguish rhetorical-role similarities that indicate relevance from superficial similarities. The dataset contains two types of query–candidate pairs: positive pairs from official relevance judgments and hard negatives that are not officially relevant but closely resemble the query. The dataset covers 807 query judgments and contains 5,184 positive pairs and 5,181 hard negatives. Positive pairs were taken directly from the official relevance judgments. Hard negatives were selected by retrieving candidates using role-based BM25 and choosing the highest-ranked candidates not included in the official relevance judgments.

We scored each query–candidate pair based on the contribution of individual rhetorical-role comparisons, such as Fact–Fact and Ratio–Ratio, to its overall similarity. The mean relevance importance score was 77.32 for positive pairs and 27.84 for hard negatives, showing a clear difference between the two groups. The role-pair analysis revealed further differences. Ratio–Ratio had the highest average importance among positive pairs (3.3796), whereas Fact–Fact was most important among hard negatives (2.3522). This suggests that matching facts alone does not necessarily indicate legal relevance. Statute–Statute had similar average importance in both groups: 1.6516 for positive pairs and 1.6690 for hard negatives.

Positive explanations often contain Ratio–Ratio, Fact–Fact, and Precedent–Precedent pairs. The mean score differences and importance rates suggest that Precedent–Ratio, Fact–Ratio, and Ratio–Ratio pairs are better at distinguishing relevant cases than Statute–Statute pairs. This analysis is descriptive: it examines the generated scores and does not directly train the final classifier. Based on these findings, we selected five role pairs—Ratio–Ratio, Fact–Fact, Precedent–Precedent, Precedent–Ratio, and Fact–Ratio—as fixed inputs to the cached ModernBERT \citep{warner-etal-2025-smarter} ranking head described in Section~\ref{subsec:role coverage}. This model is different  from the rhetorical role classifier used to assign labels text (\ref{subsec:role coverage}) and the InLegalBERT \citep{paul2023pre} LambdaRank \citep{burges2006learning} ranker described in Section \ref{subsec:lambda}. The initial and learned priors for these five role pairs changed only slightly during training, suggesting that the ranking head largely preserved the manually selected role structure. Table~\ref{tab:five-role-prior} reports the initial and learned priors for this model.

The initial data contained 5,308 positive pairs across 817 queries. Requiring usable text for at least one selected role pair reduced this to 5,179 positives across 806 queries and 4,006 positive candidates. We split the data by query, using 669 queries for training and 137 for validation. Each query group contained one positive example and seven hard or randomly selected negative examples. This resulted in 35,112 training slots and 6,320 validation slots.
After normalising the identifiers, the hard-negative file covered 806 queries, with an average of 6.43 hard negatives per query and a maximum of 51. These negative examples were useful because they shared some lexical information with the query, making them more difficult to distinguish from relevant cases.


\subsection{Cached ModernBERT Role Aware Ranking}

We used a frozen ModernBERT-base encoder \citep{warner-etal-2025-smarter} to process five selected rhetorical-role pairs: Ratio–Ratio, Fact–Fact, Precedent–Precedent, Precedent–Ratio, and Fact–Ratio. Since legal judgments often contain long sections, processing each section as a single 512-token input would leave much of the text unused. We therefore divided the text into overlapping windows of 512 tokens, with a stride of 128 tokens. Each window was converted into a 768-dimensional representation.

Encoding all windows repeatedly during training would require considerable computing resources. We therefore froze ModernBERT and stored the window representations before training the ranking head. The head used a window scorer to identify the most relevant windows and retained the top three for each role. It also used a dynamic gate and trainable role-prior weights to combine information from the selected roles when calculating the final score. This allowed the model to consider longer legal passages without repeatedly encoding the same text during training.

\begin{table}[t]
\centering
\small
\caption{Five-role prior used by the cached ModernBERT branch.}
\label{tab:five-role-prior}
\begin{tabular}{lcc}
\toprule
Role Pair & Initial Prior & Learned Prior \\
\midrule
Ratio--Ratio & 0.264564 & 0.267124 \\
Fact--Fact & 0.250755 & 0.252421 \\
Precedent--Precedent & 0.205960 & 0.203167 \\
Precedent--Ratio & 0.140259 & 0.139695 \\
Fact--Ratio & 0.138461 & 0.137593 \\
\bottomrule
\end{tabular}
\end{table}

\subsection{LambdaRank branch and official evaluation}\label{subsec:lambda}
The test set contained 237 queries and 1,727 candidate judgments, giving 409,299 possible query–candidate pairs. To reduce computation, BM25 retrieved the top 200 candidates per query for neural reranking. Of the 1,727 candidates, 1,682 (97.39\%) had valid text for at least one rhetorical-role pair. The BM25 baseline achieved Macro Recall@10 of 0.301587, Macro MRR of 0.480232, and Micro-F1@10 of 0.190804.

A parallel Phase 3 experiment kept the InLegalBERT role-pair embeddings fixed and used a LambdaRank-based ranking head with 698,614 parameters. The head was trained using an nDCG-weighted loss function. It used 5,184 positive pairs and 5,181 hard-negative pairs across 807 query groups, of which 686 were used for training and 121 for validation.

Although the network learned to produce its own ranking scores, this did not improve performance on the official retrieval set. Increasing the weight of the neural component in the ensemble led to a gradual decrease in MRR. This suggests that the cached InLegalBERT features did not improve the lexical ranking on the official test set.
We also conducted a second LambdaRank experiment using only textual information and trained the model for 10 epochs. Between epochs 9 and 10, the training loss decreased from 0.2944 to 0.2819, while the internal validation MRR remained at 0.9132. The top-10 rankings also showed very little change. These results are from internal validation and should not be confused with performance on the official retrieval set. This small improvement and the transfer failure of the previous run of LambdaRank\citep{burges2006learning} gave rise to the final design approach. The saved results from this branch are reported in Table~\ref{tab:lambdarank-results}.

\begin{table}[htbp]
\centering
\caption{Experimental data flow across pipeline stages. Counts are drawn directly from the corresponding stage's reported output; where two counts for the same nominal quantity are reported elsewhere in this paper, both are given and reconciled in the accompanying text.}
\label{tab:dataflow}
\begin{tabular}{lll}
\toprule
Stage & Queries & Candidates / Pairs \\
\midrule
Role annotated training collection & 827 & 4,319 candidates (also reported as 4,320) \\
Official test collection & 237 & 1,727 candidates; 1,727 relevant pairs \\
Phase 2: role validated inputs & 799 (valid Fact text, of 827) & 3,882 (valid Ratio text, of 4,319) \\
Phase 2: BM25 top 50 pool & --- & 39,950 pairs \\
Phase 2: after self match removal & --- & 39,271 pairs (679 removed) \\
Phase 2: final labeled supervision & 795 & 39,074 pairs \\
Phase 3: earlier audit, unfiltered & 817 & 5,308 positive pairs \\
Phase 3: after role text filtering & 806 & 5,179 positive pairs, 4,006 candidates \\
Phase 3: reported importance dataset & 807 & 5,184 positive, 5,181 hard-negative pairs \\
Phase 3: query level split (earlier branch) & 669 train / 137 validation & 35,112 train slots, 6,320 validation slots \\
Phase 3: query level split (final pipeline) & 686 train / 121 validation & --- \\
Final diversified top 600 pool (test) & 237 & 1,318 of 1,727 relevant pairs retained \\
Final entailment top 200 (test) & 237 & 1,093 of 1,727 relevant pairs retained \\
Final frozen submission (top 9) & 237 & 492 of 1,727 relevant pairs correct \\
\bottomrule
\end{tabular}
\end{table}

\begin{table}[!t]
\centering
\scriptsize
\setlength{\tabcolsep}{2.5pt}
\caption{Saved results from the earlier cached InLegalBERT LambdaRank branch.}
\label{tab:lambdarank-results}
\begin{tabular}{lcccc}
\toprule
Evaluation & MRR & MAP & nDCG@10 & Recall@10 \\
\midrule
Internal validation & 0.9711 & -- & 0.9360 & 0.8825 \\
Official text-only & 0.1414 & 0.0615 & -- & 0.4821 \\
70\% neural + 30\% BM25 & 0.1489 & 0.0654 & -- & 0.4821 \\
BM25 only (grid search) & 0.4570 & -- & -- & -- \\
\bottomrule
\end{tabular}
\end{table}

\section{Final Pipeline}\label{sec:final pipeline}

The three experimental phases helped identify the strengths and limitations of different approaches to prior case retrieval. Their findings guided the design of the final pipeline.

In Phase 1, we compared full-text BM25 and TF-IDF retrieval and explored the use of rhetorical roles for matching judgments. TF-IDF performed better than BM25 on the official test set. Among the individual role comparisons, Precedent–Precedent achieved the best MRR, MAP, and nDCG@10, while Fact–Fact gave slightly higher Recall@10. These results showed that both lexical matching and rhetorical-role information can help retrieve relevant cases, but their effectiveness depends on the type of information being compared. We retained BM25 for the later candidate-generation stages because the existing shortlist, pseudo-label, and hard-negative preparation pipelines were built around it. This choice was made to maintain continuity with the existing experiments, not because BM25 had outperformed TF-IDF.

Phase 2 explored whether comparing the Facts of a query judgment with the Ratio of the Decision of a candidate judgment could improve retrieval. We trained an InLegalBERT cross-encoder using generated support labels for this Facts-to-Ratio comparison. Although the model performed well against these generated labels, its performance did not transfer directly to the official relevance judgments. The standalone reranker performed below BM25, while combining the neural ranking with lexical retrieval gave better results. This finding showed that semantic support between two legal passages does not necessarily mean that the corresponding judgments are relevant to each other. It also highlighted the need to handle long legal passages carefully and to evaluate model selection using validation data.

Phase 3 focused on understanding how different rhetorical-role interactions contribute to relevance. The comparison of relevant pairs and hard negatives showed that some role interactions provide stronger evidence than others. Ratio–Ratio and precedent-related interactions were useful indicators among relevant cases, whereas Statute–Statute showed little difference between relevant and non-relevant pairs. In other words, two judgments may refer to the same statute without necessarily being relevant to one another. These findings suggested that the ranking model should learn the contribution of each role pair instead of treating every role match as equally useful.

Taking these findings together, we developed a four-stage pipeline. It first retrieves a broad set of candidates, then evaluates their semantic compatibility, examines their rhetorical-role relationships, and finally combines the ranking outputs.

\subsection{Role Aware Candidate Retrieval}

The first stage retrieves candidate judgments using eleven BM25 channels. One channel searches the full judgment text, while the other ten focus on specific rhetorical-role pairs: Fact–Fact, Fact–Ratio , Fact–RulingByPresentCourt, Argument–Argument, Argument–Ratio, Precedent–Precedent, Precedent–Ratio, Ratio–Ratio, Statute–Statute, Statute–Ratio

Each channel returns its top 600 candidates. Since the BM25 scores from different channels are not directly comparable, we combine their rankings using Reciprocal Rank Fusion (RRF). The score for a candidate (c) and query (q) is calculated as:
\begin{equation}
SRRF(q,c)
=
\sum_{j}
\frac{w_j}{K + r_j(q,c)},
\label{eq:srrf}
\end{equation}

where $r_j(q,c)$ represents the rank of candidate $c$ in channel $j$, $w_j$ is the corresponding channel weight, and $k=60$ is used to select candidates in the first stage. This stage combines full-text lexical retrieval with role-specific lexical matching to create a single candidate pool for the later semantic and role-aware ranking stages.

The initial ranking is then expanded to improve candidate coverage. We retain the initial top 200 candidates and add candidates retrieved by the other role-based rankings. After removing duplicates, the remaining positions are filled from the uniform ranking to form a diversified pool of up to 600 candidates per query. This step is important because the later stages can only rerank candidates that have already been retrieved. If a relevant judgment is missing from the candidate pool, the subsequent models cannot recover it.

\subsection{Sliding Window Entailment}\label{subsec:sliding window}

The second stage examines the semantic relationship between the query Facts and the candidate Ratio of the Decision using an InLegalBERT cross-encoder. The idea is to check whether the legal reasoning expressed in a candidate judgment is relevant to the factual circumstances of the query judgment.

A major challenge is that legal passages are often longer than the 512-token input limit of InLegalBERT. Simply truncating these passages could remove important information. To address this, we represent the query Facts using a 192-token head–tail strategy and divide the candidate Ratio text into overlapping windows with a 128-token overlap. During training, two windows are sampled for each query–candidate pair, while validation considers up to four windows. For each pair \(i\), let \(z_{iw}\) denote the logit for window \(w\), and let \(W_i\) be the set of windows associated with that pair. Equation~\ref{eq:log-mean-exp} defines the pair-level score using log-mean-exp aggregation:
\begin{equation}
z_i
=
\log
\left(
\sum_{w \in W_i} e^{z_{iw}}
\right)
-
\log |W_i|,
\label{eq:log-mean-exp}
\end{equation}
This aggregation allows a strongly supporting window to influence the final score without automatically giving longer judgments an advantage simply because they produce more windows.

The model uses the \texttt{law-ai/InLegalBERT} encoder. The lower transformer layers remain frozen, while the last four layers and the classification head are fine-tuned. This leaves 28,352,257 trainable parameters out of 109,483,009 total parameters. The model is trained using the generated support labels, while the official relevance labels remain the basis for evaluating retrieval performance.

After training, the selected checkpoint scores the diversified top-600 candidate pool. We then combine the neural ranking with the Stage 1 retrieval ranking using RRF. A neural-channel weight of 2.0 is selected on the validation set. The resulting top 200 candidates for each query are passed to the next stage.

\subsection{Hierarchical Role Ranker}

The third stage examines the candidates in greater detail by modelling the interactions between ten rhetorical-role pairs. The Phase 3 analysis showed that these interactions do not contribute equally to relevance. Some provide useful evidence, while others may produce misleading similarities. The role-aware ranker is therefore designed to learn both the importance and the direction of each role pair's contribution.

For each query–candidate pair, the Phase 2 fine-tuned InLegalBERT encoder processes up to two windows for each of the ten role pairs. Query roles use 160 tokens, candidate roles use 349 tokens, and candidate windows overlap by 128 tokens. The resulting embeddings are cached because the InLegalBERT encoder remains frozen during role-ranker training. Training uses query groups and combines RankNet pairwise loss, class-weighted BCE, and masked Smooth L1 loss, as defined in Equation~\ref{eq:final-objective}.

Each role-pair embedding is projected from 768 to 128 dimensions, and a role identity vector is added to retain information about the type of interaction being represented. The model then uses two prediction heads. The first predicts a signed contribution between (-1) and (1), indicating whether a role pair supports or opposes relevance. The second predicts a gate that controls how much importance the model assigns to that role pair. If a role is missing from either judgment, the corresponding pair is masked before the gate softmax is applied.

The model combines the weighted role representation with the ten predicted role contributions, the normalized Stage 2 score, and a role-coverage feature. These features are passed to a ranking network, which produces the final role-aware score. In this way, the model can learn not only which role interactions are important but also whether their presence provides positive or negative evidence for relevance.



\begin{figure}[h]
    \centering
    \includegraphics[width=1.1\textwidth]{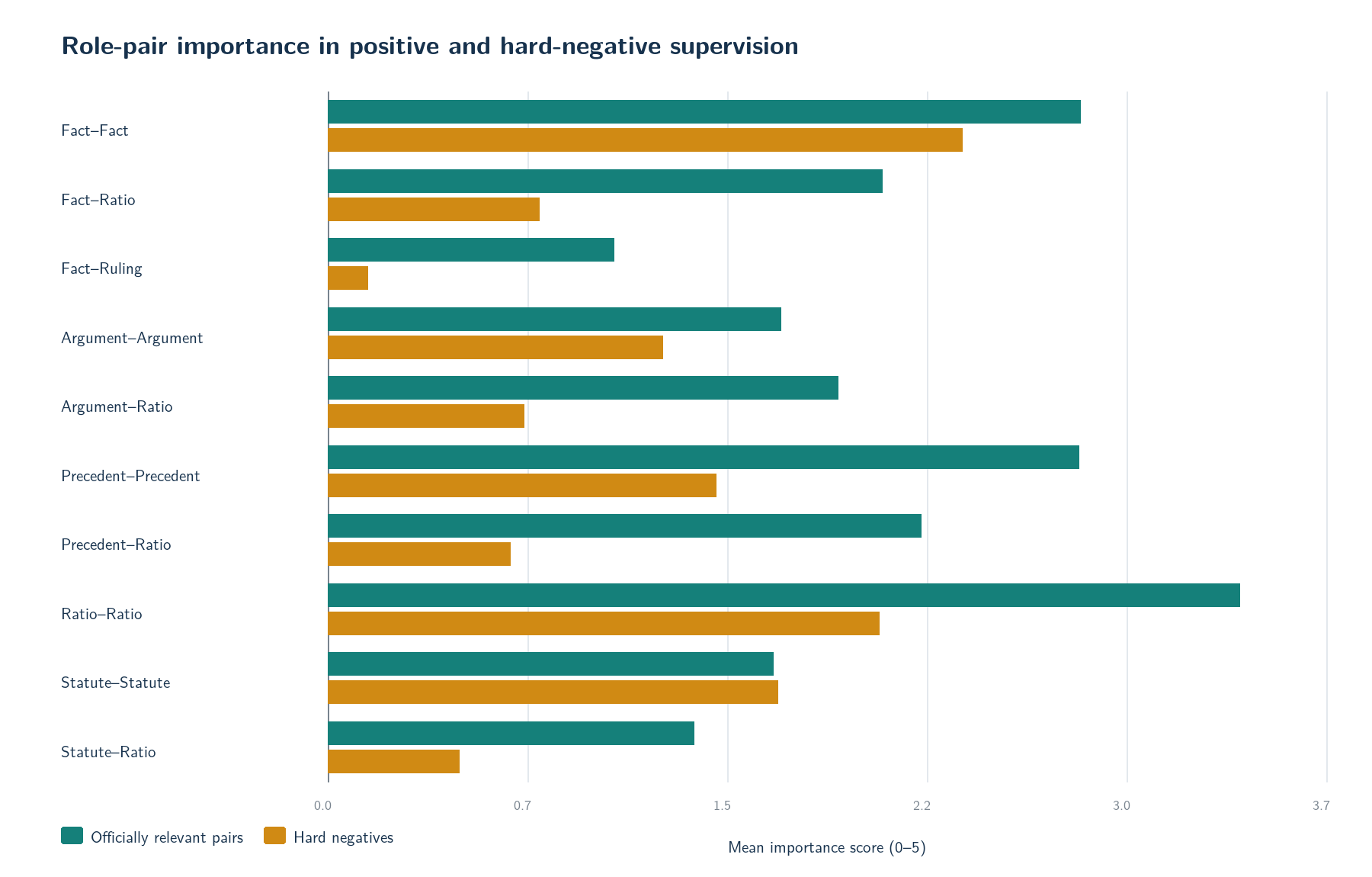}
    \caption{Mean role importance}
    \label{fig:role_importance}
\end{figure}

\begin{equation}
\mathcal{L}
=
\mathcal{L}_{\mathrm{RankNet}}
+
0.35\mathcal{L}_{\mathrm{BCE}}
+
0.40\mathcal{L}_{\mathrm{role}}.
\label{eq:final-objective}
\end{equation}


\subsection{Fusion and Evaluation}

The final stage combines the Stage 2 entailment ranking and Stage 3 hierarchical role ranking using Reciprocal Rank Fusion (RRF). We tested \(K \in \{30,60,100\}\) and different role weights relative to the entailment score. Configuration selection used a weighted criterion: 0.8 times the best Micro-F1 across cutoffs 1--20 and 0.2 times the mean Micro-F1 over the same range. The best configuration was \(K=30\), \(w_b=1\), and \(w_r=1\), with a final cutoff of 9. It achieved a validation Micro-F1 of 0.1874, MRR of 0.4563, MAP of 0.1997, and nDCG@10 of 0.2851 (Table~\ref{tab:validation-fusion}). All configuration choices were based exclusively on the validation set, and the final settings were fixed before testing. Official test labels were not used for training or model and configuration selection.

The submission contained 237 query IDs, each with nine candidates. We checked for duplicate query IDs, self-retrieved cases, unknown candidates, and entries outside the allowed pool. Following official evaluation, a technical audit was conducted without changing model parameters, rankings, fusion weights, or cutoffs.

\begin{figure}[t]
    \centering
    \includegraphics[width=1.1\textwidth]{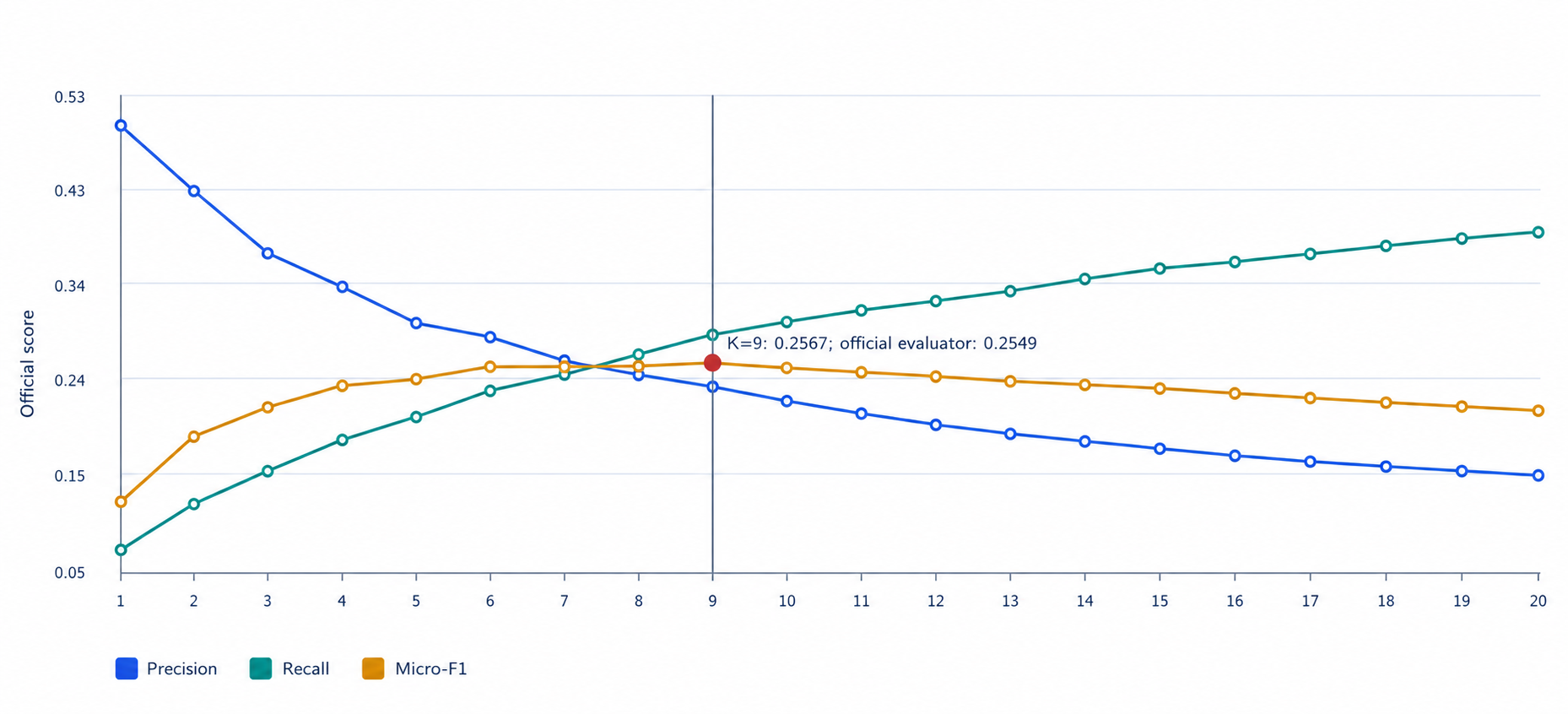}
    \caption{Post-hoc cutoff curve}
    \label{fig:final}
\end{figure}

Figure~\ref{fig:final} shows the post-hoc Micro-F1 across different cutoffs. The highlighted value of 0.2567 was calculated before the denominator correction and is not an official result. The corrected evaluator reports an official Micro-F1 of 0.254922 at cutoff 9 (Table~\ref{tab:official-evaluation}).

The diversified top-600 pool contains 1,318 of the 1,727 official relevant pairs, of which 1,093 remain in the top 200 after entailment reranking. At the final cutoff, entailment ranking retrieves 408 relevant pairs, compared with 492 for final fusion. The query-level audit records 199 successful queries, 21 final-ranking failures, 10 retrieval failures, and 7 pruning failures. Compared with the Section 4 ranking, role fusion improves 85 queries, leaves 116 unchanged, and worsens 36. This indicates that its main benefit comes from the final ranking stage.

Analysis of mean predicted role contributions shows the largest positive differences for Fact--RulingByPresentCourt, Precedent--Precedent, Ratio--Ratio, Fact--Fact, and Precedent--Ratio interactions. Statute-only interactions show a negative difference, supporting the earlier finding that shared statutory language may indicate lexical similarity without establishing legal relevance. Overall, fact- and precedent-related interactions contribute more to ranking relevant cases than statute-only interactions.


The final ranking combines the Phase 2 entailment rank and the role aware rank using reciprocal rank fusion. For a query q and candidate c, the final score is defined as in Equation \ref{eq:final-fusion}

\begin{equation}
S_{\mathrm{final}}(q,c)
=
\frac{w_b}{K+r_b(q,c)}
+
\frac{w_r}{K+r_r(q,c)},
\label{eq:final-fusion}
\end{equation}


\begin{table}[t]
\centering
\scriptsize
\setlength{\tabcolsep}{10.5pt}
\caption{Leading validation fusion configurations.}
\label{tab:validation-fusion}
\begin{tabular}{lccccc}
\toprule
Configuration & Best $k$ & Micro F1 & MRR & MAP & nDCG@10 \\
\midrule
$k=30$, base 1, role 1
& 9 & 0.1874 & 0.4563 & 0.1997 & 0.2851 \\

$k=60$, base 1, role 1
& 8 & 0.1867 & 0.4526 & 0.2010 & 0.2879 \\

$k=100$, base 1, role 1
& 6 & 0.1799 & 0.4483 & 0.1982 & 0.2829 \\

$k=100$, base 1, role 2
& 8 & 0.1779 & 0.4120 & 0.1744 & 0.2586 \\

Entailment(stage 2), $k=30$
& 9 & 0.1349 & 0.3331 & 0.1494 & 0.2028 \\
\bottomrule
\end{tabular}
\end{table}



\begin{table}[t]
\centering
\small
\caption{Corrected official evaluation of the frozen top 9 submission.}
\label{tab:official-evaluation}
\begin{tabular}{lc|lc}
\toprule
Metric & Score & Metric & Score \\
\midrule
Micro Precision@9 & 0.230660 & MRR & 0.608099 \\
Micro Recall@9 & 0.284887 & MAP & 0.315868 \\
Micro F1@9 & 0.254922 & nDCG@10 & 0.418657 \\
\bottomrule
\end{tabular}
\end{table}

\begin{figure}[t]
    \centering
    \includegraphics[width=1.1\textwidth]{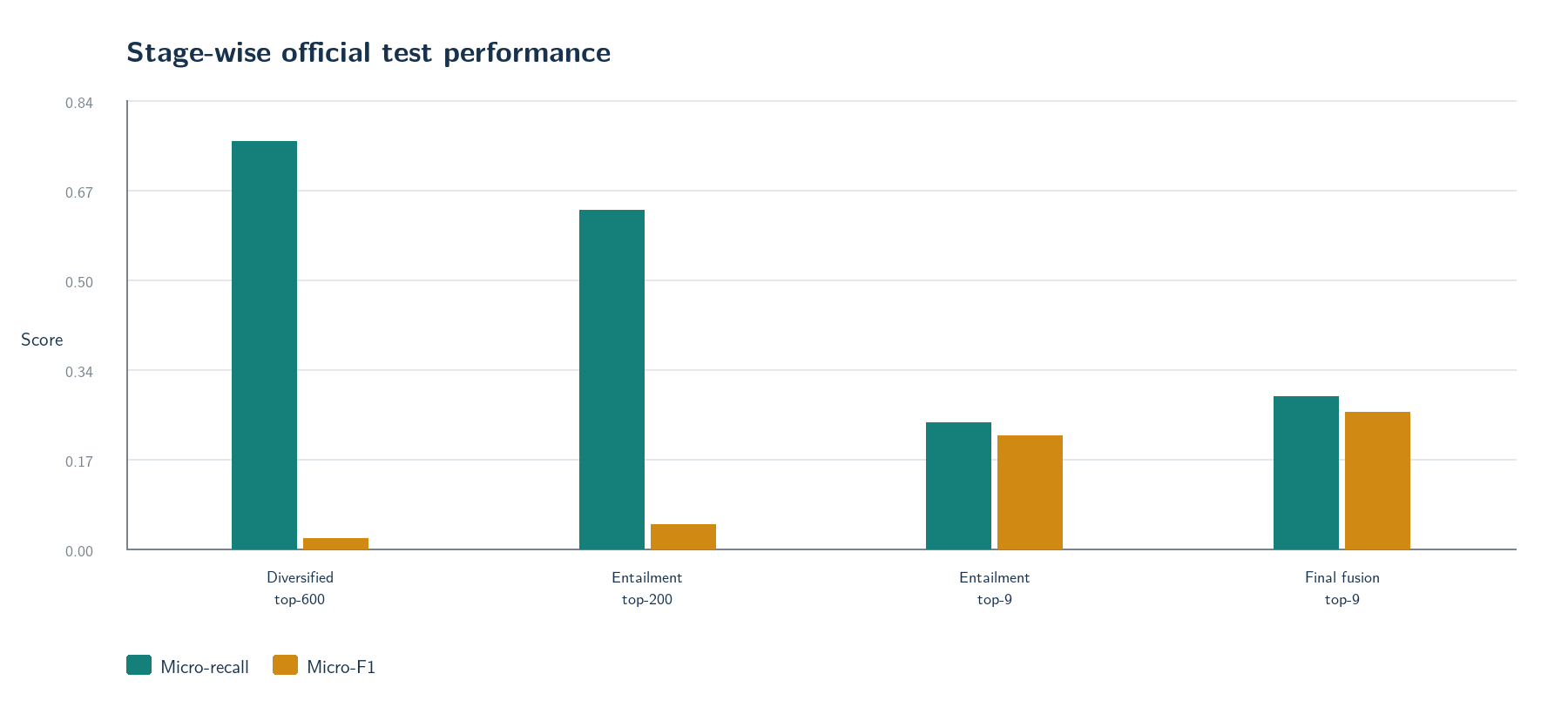}
    \caption{Recall and Micro F1 at successive stages of the pipeline}
    \label{fig:stage_perf}
\end{figure}


\section{Results and Discussions}\label{sec:results}
The results in this study are divided into three categories: official, validation, and internal. Only the official results, calculated on the 237-query test set using the official relevance judgments, are used to evaluate the performance of the proposed method. Tables \ref{tab:phase1_results}, \ref{tab:bm25_shortlist}, and \ref{tab:stage_wise_results} present the results of the methods evaluated under these three categories.


The baseline study showed that TF-IDF with cosine normalisation outperformed BM25 at all tested retrieval depths, highlighting the value of lexical weighting in legal retrieval. However, BM25 was retained to match the existing labelling pipeline. Among the rhetorical-role pairs, Precedent–Precedent and Fact–Fact were useful, while Issue–Ratio was less effective. Combining role pairs performed better than using any single pair, suggesting that different roles provide complementary signals. Still, role-based retrieval did not outperform full-text lexical retrieval. The entailment study showed that strong results on generated labels do not always lead to better official retrieval results. The Facts-to-Ratio reranker performed well on Gemini-generated labels but fell below BM25 in the official evaluation. Combining semantic and lexical scores improved the official classification measure compared with either measure alone, suggesting that semantic features can complement lexical matching. The hierarchical sliding-window method provided no further improvement over this combination. However, the combination weight was selected using test data, so the improvement cannot be treated as an independent evaluation.

\begin{figure}[H]
    \centering
    \resizebox{0.9\linewidth}{!}{\includegraphics{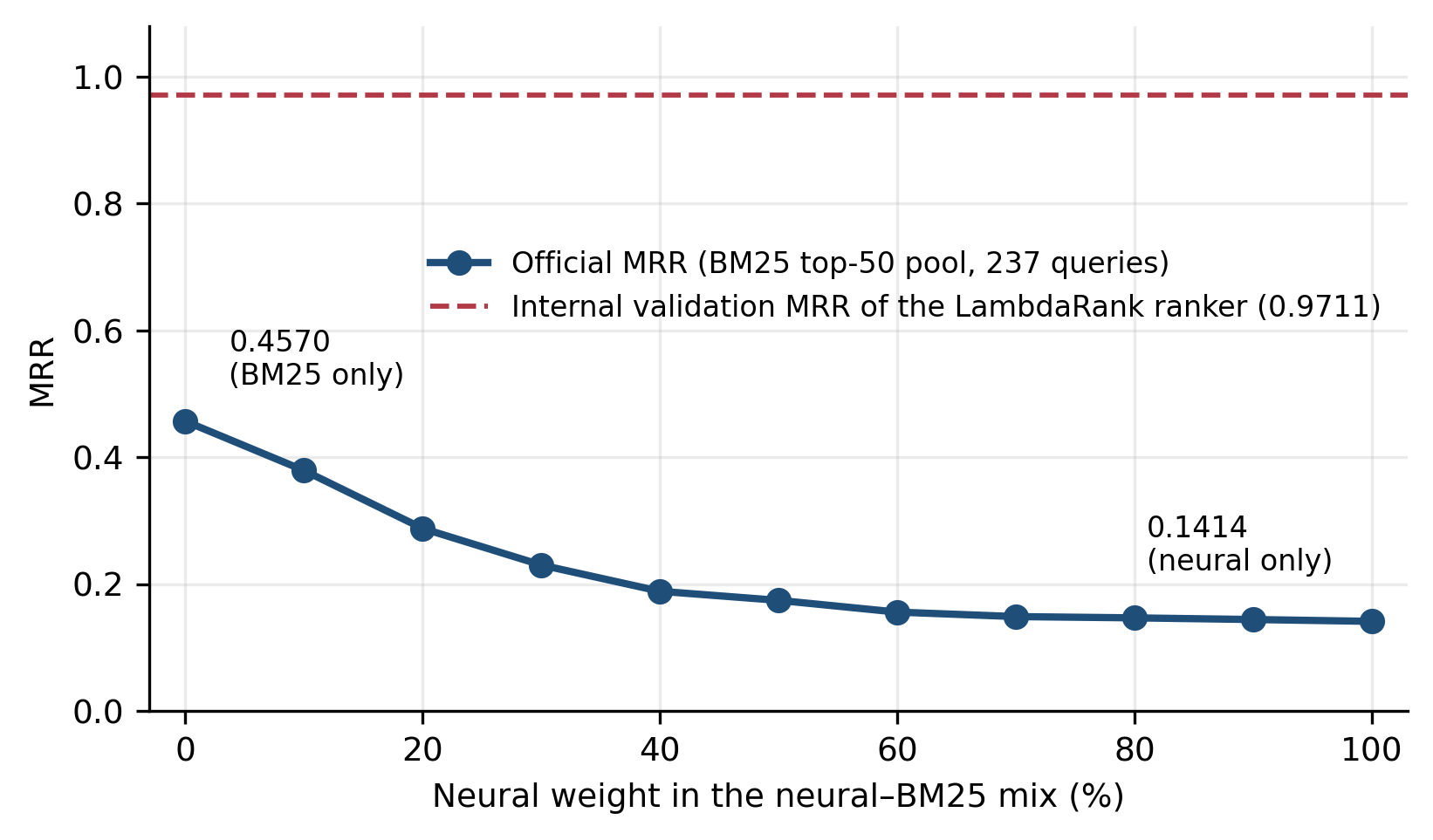}}
    \caption{Effect of neural-score weighting on official-test MRR for neural–BM25 fusion, compared with the BM25-only baseline}
    \label{fig:sweep}
\end{figure}

The transfer study showed that strong internal results did not always translate to good performance on the official test set. LambdaRank performed well on a restricted candidate set, but its performance dropped during official evaluation. Figure~\ref{fig:sweep} also shows that increasing the neural score weight reduced performance, with lexical ranking performing best. However, since the weight sweep used the official test set, these findings are diagnostic and do not provide reliable evidence of generalisation. The ModernBERT head was tested only internally, so its performance on the official test set remains unknown.

\begin{figure}[!ht]
    \centering
    \includegraphics[width=0.75\linewidth]{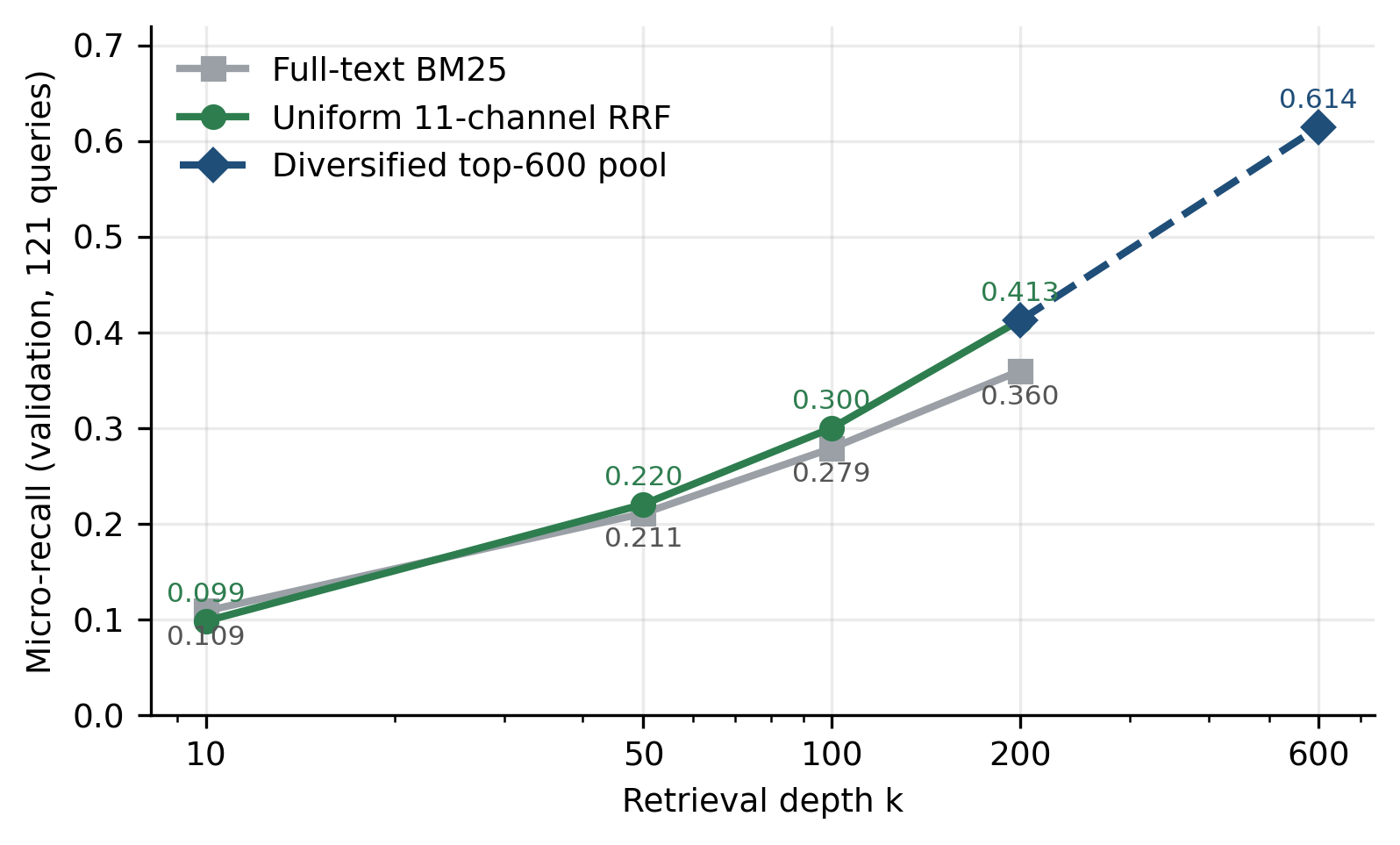}
    \caption{Validation Micro-Recall at increasing retrieval depths for full-text BM25, uniform eleven-channel reciprocal rank fusion, and the diversified candidate pool}
    \label{fig:stage_recall}
\end{figure}

The final pipeline was selected using the validation split and kept unchanged for the official test evaluation (Table~\ref{tab:stage_wise_results}). An error in the evaluator's denominator was corrected without changing the models, rankings, weights, or cutoff. The stage-wise results highlight the importance of candidate coverage: later stages cannot recover relevant cases missed during initial retrieval. Role-based channels found more relevant cases at greater retrieval depths but did not improve the top-ranked results (Figure~\ref{fig:stage_recall}). This suggests that their main benefit is expanding candidate coverage rather than improving the initial ranking. Later filtering and ranking stages reduced the candidate pool but also removed some relevant cases.

\begin{figure}[H]
    \centering
    \resizebox{0.9\linewidth}{!}{
        \includegraphics{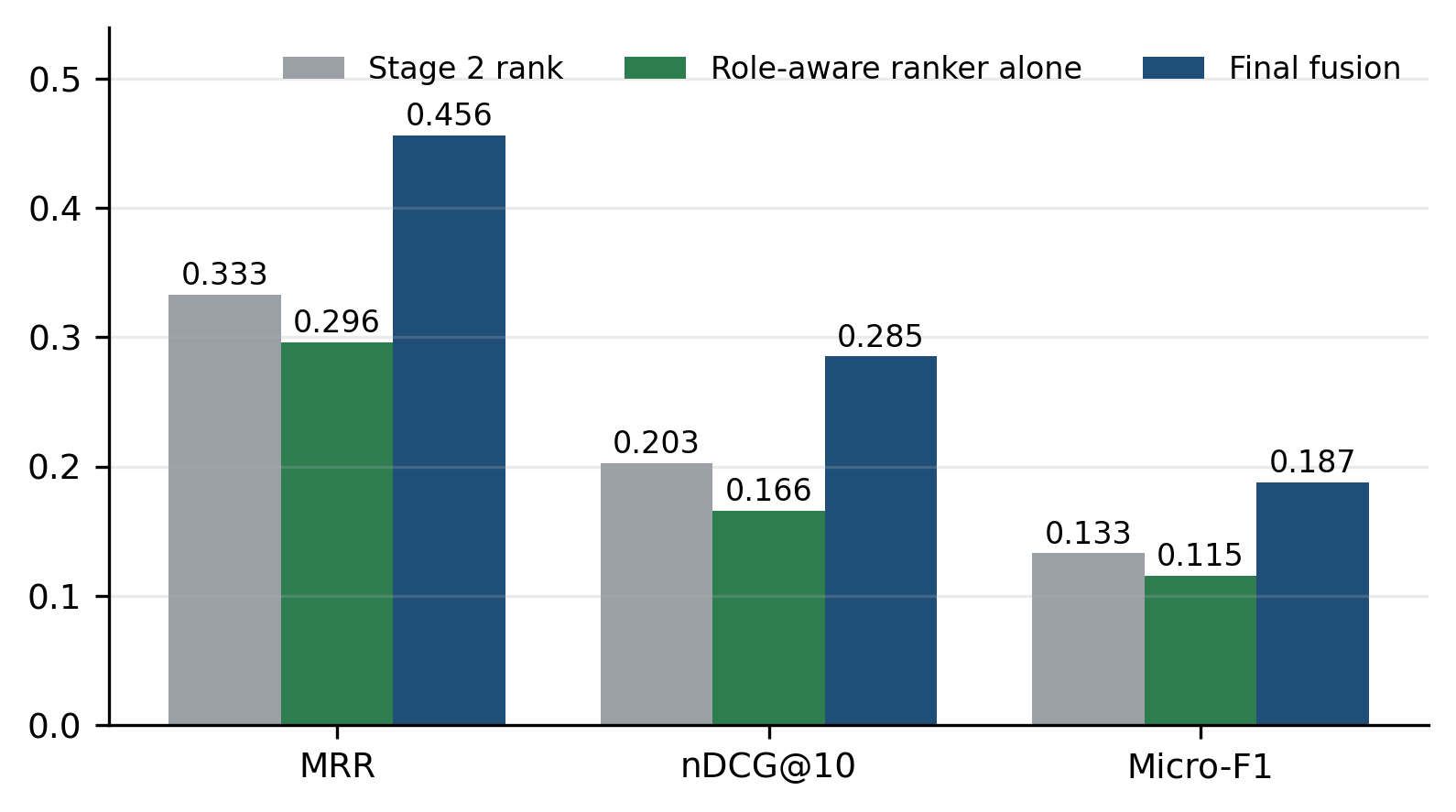}
    }
    \caption{Validation comparison of entailment-based ranking, standalone role-aware ranking, and final reciprocal rank fusion}
    \label{fig:role_fusion}
\end{figure}

The role-aware ranker performed worse than the entailment-based ranker on the validation set (Figure \ref{fig:role_fusion}). However, combining the two methods gave better results than using either one alone. The combined system also performed better than the preceding ranking stage under official evaluation (Table \ref{tab:stage_wise_results}). This shows that combining different ranking signals can be more useful than relying on the role-aware ranker alone.
The improvement was not the same for every query. Fusion improved the rankings for 85 queries, made no difference for 116, and worsened the rankings for 36. The remaining errors occurred at different stages, including retrieving relevant cases, filtering candidates, and ranking the final results. Addressing these issues at each stage may further improve the system.

The results from the post hoc analysis indicate that pairs of role types based on facts, precedent, and outcomes are informative, whereas statute only pairs were not, and had an inverse relationship. This implies that cases with the same legal provisions could be generally similar without necessarily constituting a relevant precedent. Also, the fact that cases have the same set of facts is not enough to establish relevance since even hard negatives can have the same set of facts as the query.


Table \ref{tab:ilpcr_comparison} compares the proposed pipeline with TraceRetriever \citep{nigam2025segment} on the IL-PCR benchmark, with both methods evaluated at a cutoff of 5. The proposed pipeline achieves higher Precision, F1 and MRR. In contrast, TraceRetriever achieves slightly higher Recall and MAP. This shows that the proposed pipeline performs better in terms of precision, F1, and the ranking of relevant cases, while TraceRetriever retrieves a slightly larger proportion of relevant cases and achieves a higher MAP. However, the results should be viewed with some caution, as the two methods may differ in their query sets and evaluation procedures. Therefore, this comparison should not be taken as evidence that one method is better in all settings. Table \ref{tab:ablation_study} summarises the experimental results across all three phases, showing how lexical, semantic, and rhetorical-role-based retrieval methods performed under different evaluation settings and how they contributed to the development of the final retrieval pipeline.

The experiments also highlight three important findings. First, rhetorical roles can provide useful signals for prior case retrieval, but their contribution varies across role pairs. Second, these signals are more useful when combined with lexical and semantic information rather than used independently. Third, the choice of evaluation setting can have a noticeable effect on the results. In particular, performance on generated labels and restricted candidate sets did not always reflect performance on the official test set. Overall, the findings suggest that combining lexical, semantic, and rhetorical-role information can improve prior case retrieval under the evaluated setting. Further experiments on other datasets and legal domains are needed to understand whether these findings hold more broadly.






\begin{table}[!htbp]
\centering
\small
\setlength{\tabcolsep}{4pt}
\renewcommand{\arraystretch}{1.1}
\caption{Phase 1 (Baseline Study) results for full-text and role-based retrieval. TF-IDF outperforms BM25 at the tested retrieval depths. Among individual rhetorical role pairs, Precedent--Precedent is the strongest, while Issue--Ratio provides little retrieval benefit.}
\label{tab:phase1_results}

\noindent\makebox[\linewidth][l]{\textbf{(a) Full-text retrieval}}\par\smallskip
\noindent\begin{tabular*}{\linewidth}{@{}l@{\extracolsep{\fill}}rrrrrr@{}}
\toprule
Method & $k$ & Precision & Recall & MRR & MAP & nDCG \\
\midrule
BM25   & 10 & 0.1239 & 0.2194 & 0.4252 & 0.1492 & 0.2444 \\
TF-IDF & 10 & 0.3154 & 0.5700 & 0.7385 & 0.4376 & 0.5932 \\
BM25   & 50 & 0.0428 & 0.3549 & 0.4339 & 0.1642 & 0.2765 \\
TF-IDF & 50 & 0.0958 & 0.7375 & 0.7401 & 0.4848 & 0.6352 \\
\bottomrule
\end{tabular*}

\vspace{1em}

\noindent\makebox[\linewidth][l]{\textbf{(b) Role-based retrieval}}\par\smallskip
\noindent\begin{tabular*}{\linewidth}{@{}l@{\extracolsep{\fill}}rrrrrr@{}}
\toprule
Method & R@1 & R@5 & R@10 & MRR & MAP & nDCG@10 \\
\midrule
Fact--Fact                & 0.0600 & 0.1517 & 0.2021 & 0.3446 & 0.1159 & 0.2007 \\
Fact--Ratio               & 0.0138 & 0.0725 & 0.1078 & 0.1948 & 0.0504 & 0.1074 \\
Ratio--Ratio              & 0.0435 & 0.1156 & 0.1649 & 0.3098 & 0.0941 & 0.1759 \\
Precedent--Precedent      & 0.0618 & 0.1553 & 0.2013 & 0.3572 & 0.1258 & 0.2092 \\
Issue--Ratio              & 0.0011 & 0.0034 & 0.0054 & 0.0151 & 0.0020 & 0.0054 \\
Logistic-regression comb. & 0.0715 & 0.1923 & 0.2641 & 0.4329 & 0.1568 & 0.2647 \\
\bottomrule
\end{tabular*}
\end{table}

\begin{figure}[H]
    \centering
    \resizebox{0.9\linewidth}{!}{\includegraphics{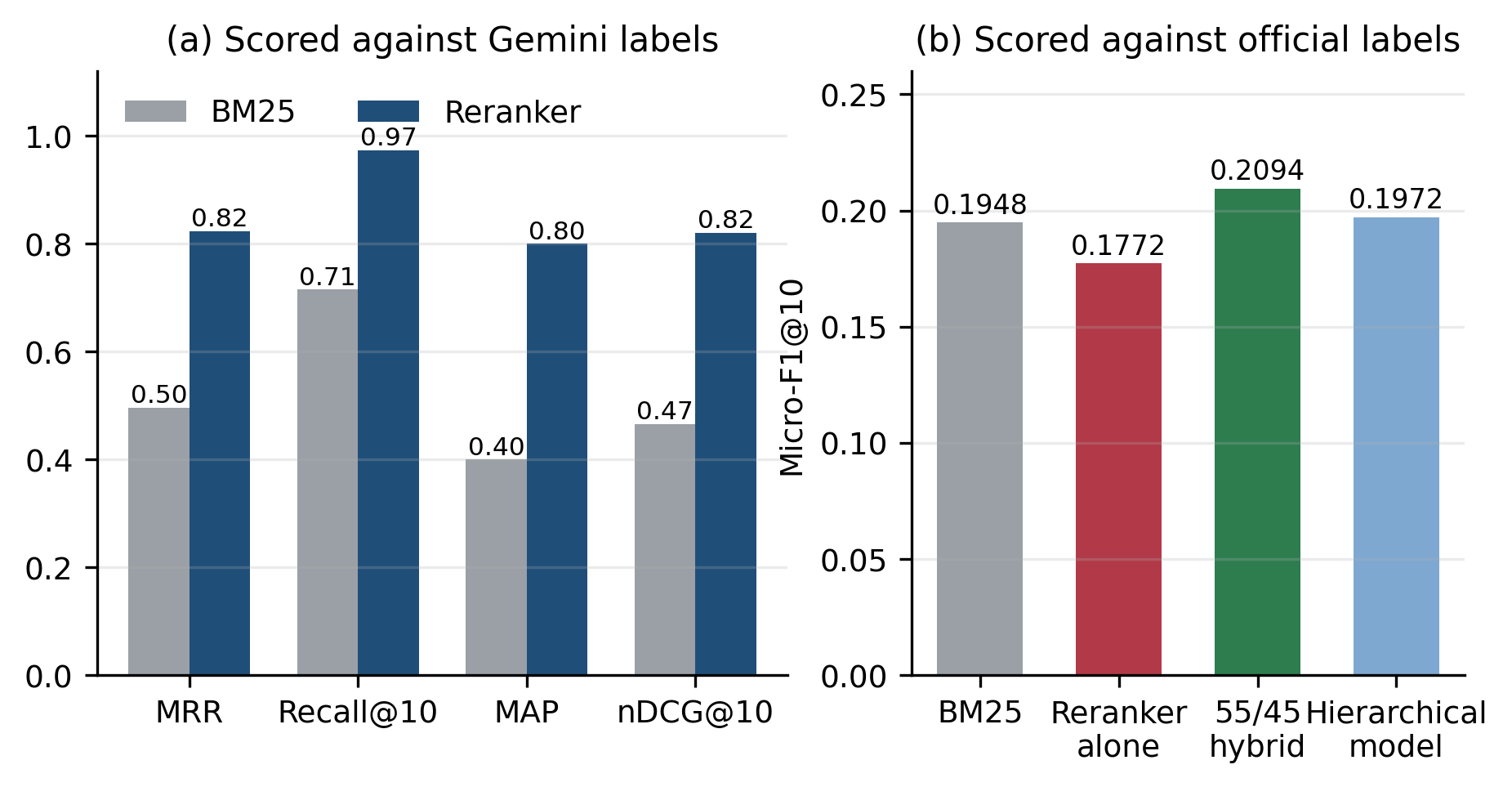}}
    \caption{Comparison of lexical retrieval, semantic reranking, and role-aware ranking configurations using their respective evaluation metrics}
    \label{fig:transfer}
\end{figure}

Figure \ref{fig:transfer} show that performance on generated relevance labels does not always reflect performance under the official evaluation. In our experiments, combining lexical and semantic scores performed better than using the semantic reranker alone. This suggests that semantic matching can improve retrieval when used alongside lexical matching. However, this improvement should be interpreted with caution, as the combination weights were selected using the official test set, so the results do not provide an independent estimate of performance on unseen data.

\begin{table}[!h]
\centering
\small                                    
\setlength{\tabcolsep}{4pt}               
\renewcommand{\arraystretch}{1.1}         
\caption{Official scores: BM25 shortlist evaluated on the official test collection. As the search depth ($K$) increases from $1$ to $200$, the system finds more relevant cases (raising the hit rate and recall).  }
\label{tab:bm25_shortlist}
\begin{tabular*}{\linewidth}{@{}l@{\extracolsep{\fill}}cccccc@{}}   
\toprule
$K$ & Hit rate & Recall & Precision & MRR & DCG & Micro-F1 \\
\midrule
1   & 0.3846 & 0.0901 & 0.3846 & 0.3846 & 0.3846 & 0.0918 \\
10  & 0.7222 & 0.3016 & 0.1658 & 0.4802 & 0.3067 & 0.1908 \\
100 & 0.8675 & 0.5440 & 0.0354 & 0.4858 & 0.3733 & 0.0659 \\
200 & 0.9145 & 0.6384 & 0.0216 & 0.4862 & 0.3975 & 0.0418 \\
\bottomrule
\end{tabular*}                            
\end{table}

The results from figure \ref{fig:sweep} show that good performance during internal testing does not necessarily lead to better results under the official evaluation. In our experiments, giving more weight to the neural score did not improve the official results. This highlights the need to select models using validation data that is separate from the training queries and to evaluate all methods using the same procedure. Since the weight settings were tested on the official test set, these findings should be treated as observations from the experiment rather than independent evidence that the method will perform well on unseen data.

The results from figure \ref{fig:role_fusion} show that combining entailment-based ranking with role-aware ranking improves the final list of retrieved cases compared with the previous ranking stage. The role-aware ranker does not perform better on its own, but it provides additional information that helps improve the final results when combined with the other ranking method. However, this improvement does not occur for every query, so the overall gain should not be taken to mean that every query benefits from the combination.

Figure \ref{fig:stage_recall} shows that role-based retrieval can help find more relevant cases when the system searches a larger pool of candidates, even if it does not improve the highest-ranked results. This suggests that role-based channels are useful for expanding the candidate pool before applying more selective ranking methods. The analysis also highlights the importance of candidate recall: if a relevant case is missed during the initial retrieval stage, later stages cannot recover it.

\clearpage


\begin{table}[!htbp]
\centering
\small                                    
\setlength{\tabcolsep}{4pt}               
\renewcommand{\arraystretch}{1.1}         
\caption{Stage-wise official results (237 test queries)}
\label{tab:stage_wise_results}
\begin{tabular*}{\linewidth}{@{}l@{\extracolsep{\fill}}ccccc@{}}   
\toprule
\textbf{Stage} & \textbf{Relevant pairs} & \textbf{Predictions} & \textbf{Precision} & \textbf{Recall} & \textbf{Micro-F1} \\
\midrule
Diversified top-600 pool & 1,318 & 142,200 & 0.0093 & 0.7632 & 0.0183 \\
Stage 2 reranked top-200 & 1,093 & 47,400 & 0.0231 & 0.6329 & 0.0445 \\
Stage 2 rank, top-9 & 408 & 2,133 & 0.1913 & 0.2362 & 0.2114 \\
Final fusion, top-9 & 492 & 2,133 & 0.2307 & 0.2849 & 0.2549 \\
\bottomrule
\end{tabular*}                           
\end{table}

\section{Comparison with existing Prior Case Retrieval Methods}
To evaluate the proposed framework in relation to existing approaches, we compare its results with two studies that have reported experimental findings on the IL-PCR dataset, \citep{joshi-etal-2023-u} introduced U-CREAT, an unsupervised event-based retrieval framework that extracts events from legal judgments to identify relevant prior cases. Their experiments showed that the approach improved retrieval performance over BM25 on the IL-PCR and COLIEE datasets. Their best-performing configuration, which applies quad-gram BM25 to event-filtered documents, achieved a Precision of 0.4012, Recall of 0.3822, and F1-score of 0.3915 at \(k=7\). Similarly \citep{premasiri2025llm} explored large language model-based embedding methods for legal case retrieval across four datasets, including IL-PCR. Their best-performing model, SFR-Embedding-2 R, achieved a MAP of 0.47 and an F-score of 0.34 at \(k=5\) on IL-PCR, outperforming BM25 and the supervised LegalBERT baseline in their experimental setting. However, the study did not report Precision, Recall, or MRR for this model on IL-PCR.

Our work takes a different approach by combining lexical, semantic, and rhetorical-role information in a multi-stage retrieval framework. It brings together eleven BM25 retrieval channels, a sliding-window InLegalBERT cross-encoder, a directed rhetorical-role-pair ranker, and validation-based Reciprocal Rank Fusion (RRF). For a more consistent comparison with the results reported at (k=5), the cutoff commonly adopted in prior retrieval studies, we also evaluate our proposed framework at the same cutoff. On the official IL-PCR test set, the proposed framework achieved a corrected Micro-F1 of 0.254922, an MRR of 0.608099, a MAP of 0.315868, and an nDCG@10 of 0.418657. We compare these results with the findings of the two studies to examine how the proposed approach relates to existing event-based and embedding-based retrieval methods.

\section{Evaluation Metrics and Experimental Considerations}
\begin{itemize}

\item P@k, R@k, and Micro-F1@k denote precision, recall, and micro-F1 at cutoff $k$, respectively. R@1, R@5, and R@10 denote recall at the corresponding ranks. MRR, MAP, and nDCG@10 denote mean reciprocal rank, mean average precision, and normalized discounted cumulative gain at rank 10, respectively.

\item A dash (--) indicates that the corresponding metric was not reported for that experiment. All values are reproduced from the respective experimental tables and have not been recalculated.

\item The evaluation split used for the Phase I logistic-regression experiment should be distinguished from the evaluation setting of the lexical baselines. The Phase II Gemini-generated-label results are reported separately because they use automatically generated support labels rather than the official IL-PCR relevance labels.

\item The Phase II 55:45 hybrid configuration was selected using the official test set. Therefore, its result should be interpreted as an exploratory finding rather than an independent estimate of test-set generalization.

\item The Phase III BM25 shortlist experiment uses a different candidate-pool setting from the Phase II BM25 baseline. Its reported MRR is macro-averaged, whereas Micro-F1 is micro-averaged.

\item The final-stage rows represent successive stages of the retrieval pipeline, each operating at a different candidate depth. They should therefore be interpreted as a retrieval funnel rather than as directly comparable top-$k$ ablation configurations.

\item Finally, the RRF configuration was selected using validation experiments before evaluation on the official test set. The final official-test results are reported at cutoff $k=9$.
\end{itemize}

\section{Conclusion}\label{sec:conclusions}
In this research, we developed and evaluated a complete prior case retrieval pipeline for Indian legal documents, covering both lexical and semantic retrieval methods with rhetorical-role information. The initial experiments showed the importance of exact lexical matching and demonstrated that combining information from different rhetorical roles can improve retrieval. The initial Facts-to-Ratio-of-the-Decision reranking model performed well when trained and evaluated using generated labels, but its performance decreased when tested with official relevance judgments. This led us to introduce sliding-window processing, query-based data splitting, and model selection based only on validation results.

 This study reveals that rhetorical-role information constitutes an interesting complementary signal for prior-case retrieval in the Indian legal domain. The results of the Phase 1 experiments indicate that different role-pairs play a distinct role in the case retrieval process and that using several role-pairs at once leads to better performance compared to a single one. Nevertheless, the findings of this study do not confirm the assumption that rhetorical role retrieval performs better than the full-text lexical retrieval. Therefore, the final retrieval pipeline uses both full-text and role-based lexical retrieval channels together with semantic entailment reranking and role-based ranking. Improvement of the second stage of the pipeline in the official evaluation supports the idea that rhetorical roles information could become a complementary signal for retrieval in multi-stage system.

The proposed system consists of eleven BM25-based retrieval channels for candidate selection, a partially fine-tuned InLegalBERT cross-encoder for legal entailment, a signed contribution ranker based on ten rhetorical-role pairs, and a Reciprocal Rank Fusion (RRF) model with settings selected on the validation set. On the official evaluation set, the final top-9 results achieved a corrected Micro-F1 of 0.254922, an MRR of 0.608099, a MAP of 0.315868, and an nDCG@10 of 0.418657. The addition of rhetorical-role information through fusion improved Micro-F1 by 20.6\% compared with the previous ranking stage. It also increased successful query coverage from 76.37\% to 83.97\%.

\begin{table*}[t]
\centering
\caption{Comparison of prior case retrieval methods on the IL-PCR benchmark. }
\label{tab:ilpcr_comparison}
\small
\setlength{\tabcolsep}{3pt}
\begin{tabular}{@{}lcccccc@{}}
\toprule
\textbf{Method} &
\textbf{P} &
\textbf{R} &
\textbf{F1} &
\textbf{MAP} &
\textbf{MRR} &
\textbf{Cutoff} \\
\midrule
TraceRetriever(2025)
& 0.1843 & 0.2088 & 0.1636 & 0.3783 & 0.3924 & 5 \\
\textbf{Proposed Final Pipeline}
& \textbf{0.292827}
& \textbf{0.200926}
& \textbf{0.238324}
& \textbf{0.327444}
& \textbf{ 0.589522}
& \textbf{5} \\
\bottomrule
\end{tabular}
\end{table*}

\begin{table*}[htbp]
\centering
\caption{Phase-wise experimental comparison of lexical, semantic, and
rhetorical-role-based retrieval configurations.}
\label{tab:ablation_study}

\small
\setlength{\tabcolsep}{1.8pt}
\renewcommand{\arraystretch}{1.15}

\begin{threeparttable}

\resizebox{0.96\textwidth}{!}{%
\begin{tabular}{llclccccccccc}
\toprule
\textbf{Phase} &
\textbf{Method} &
\textbf{Evaluation setting} &
\textbf{Cutoff} &
\textbf{P@k} &
\textbf{R@k} &
\textbf{Micro-F1@k} &
\textbf{R@1} &
\textbf{R@5} &
\textbf{R@10} &
\textbf{MRR} &
\textbf{MAP} &
\textbf{nDCG@10} \\
\midrule


\multirow{10}{*}{I}
& BM25 & IL-PCR
& 10 & 0.1239 & 0.2194 & -- & -- & -- & -- & 0.4252 & 0.1492 & 0.2444 \\

& BM25 & IL-PCR
& 50 & -- & 0.3549 & -- & -- & -- & -- & -- & 0.1642 & -- \\

& TF-IDF & IL-PCR
& 10 & 0.3154 & 0.5700 & -- & -- & -- & -- & 0.7385 & 0.4376 & 0.5932 \\

& TF-IDF & IL-PCR
& 50 & -- & 0.7375 & -- & -- & -- & -- & -- & 0.4848 & -- \\

& Fact--Fact & Role-pair retrieval
& -- & -- & -- & -- & 0.0600 & 0.1517 & 0.2021 & 0.3446 & 0.1159 & 0.2007 \\

& Fact--Ratio & Role-pair retrieval
& -- & -- & -- & -- & 0.0138 & 0.0725 & 0.1078 & 0.1948 & 0.0504 & 0.1074 \\

& Ratio--Ratio & Role-pair retrieval
& -- & -- & -- & -- & 0.0435 & 0.1156 & 0.1649 & 0.3098 & 0.0941 & 0.1759 \\

& Precedent--Precedent & Role-pair retrieval
& -- & -- & -- & -- & 0.0618 & 0.1553 & 0.2013 & 0.3572 & 0.1258 & 0.2092 \\

& Issue--Ratio & Role-pair retrieval
& -- & -- & -- & -- & 0.0011 & 0.0034 & 0.0054 & 0.0151 & 0.0020 & 0.0054 \\

& Logistic regression & Role-pair combination
& -- & -- & -- & -- & 0.0715 & 0.1923 & 0.2641 & 0.4329 & 0.1568 & 0.2647 \\

\midrule


\multirow{5}{*}{II}
& BM25 & Official evaluation
& 10 & -- & -- & 0.194777 & -- & -- & -- & -- & -- & -- \\

& InLegalBERT & Official evaluation
& 10 & -- & -- & 0.177 & -- & -- & -- & -- & -- & -- \\

& BM25 + InLegalBERT & 55:45 hybrid
& 10 & 0.181013 & 0.248408 & 0.209422 & -- & -- & -- & -- & -- & -- \\

& Hierarchical sliding-window & Official evaluation
& 10 & 0.1705 & 0.2339 & 0.1972 & -- & -- & -- & -- & -- & -- \\

& InLegalBERT & Gemini-generated labels
& 10 & -- & 0.973231 & -- & -- & -- & -- & 0.822895 & 0.799915 & 0.819971 \\

\midrule


\multirow{6}{*}{III}
& BM25 & Top-200 shortlist
& 10 & 0.1658 & 0.3016 & 0.1908 & -- & -- & -- & 0.4802 & -- & -- \\

& Stage 2 entailment ranking & Validation (121 queries)
& 9 & -- & -- & 0.1349 & -- & -- & -- & 0.3331 & 0.1494 & 0.2028 \\

& RRF ($K=30$, base=1, role=1) & Validation (121 queries)
& 9 & -- & -- & 0.1874 & -- & -- & -- & 0.4563 & 0.1997 & 0.2851 \\

& RRF ($K=60$, base=1, role=1) & Validation (121 queries)
& 8 & -- & -- & 0.1867 & -- & -- & -- & 0.4526 & 0.2010 & 0.2879 \\

& RRF ($K=100$, base=1, role=1) & Validation (121 queries)
& 6 & -- & -- & 0.1799 & -- & -- & -- & 0.4483 & 0.1982 & 0.2829 \\

& RRF ($K=100$, base=1, role=2) & Validation (121 queries)
& 8 & -- & -- & 0.1779 & -- & -- & -- & 0.4120 & 0.1744 & 0.2586 \\

\midrule


\multirow{4}{*}{Final}
& Diversified candidate pool & Official test (237 queries)
& 600 & 0.0093 & 0.7632 & 0.0183 & -- & -- & -- & -- & -- & -- \\

& Stage 2 reranked candidates & Official test (237 queries)
& 200 & 0.0231 & 0.6329 & 0.0445 & -- & -- & -- & -- & -- & -- \\

& Stage 2 entailment ranking & Official test (237 queries)
& 9 & 0.1913 & 0.2362 & 0.2114 & -- & -- & -- & -- & -- & -- \\

& Final RRF fusion & Official test (237 queries)
& 9 & 0.2307 & 0.2849 & 0.2549 & -- & -- & -- & 0.6081 & 0.3159 & 0.4187 \\

\bottomrule
\end{tabular}%
}

\end{threeparttable}
\end{table*}

\section{Limitations}\label{sec:limitations}
This work is based on the analysis of judgments from the Indian Supreme Court and does not necessarily hold true for any other judicial body. This is because the findings are dependent on the presence of annotated relevance judgments and rhetorical role labels. Rhetorical role coverage is uneven across judgments, with some roles unavailable for a substantial portion of the collection and processing long legal sections remains computationally demanding task as well.










\begin{appendices}

\section{Evaluation Metrics}\label{sec:evaluation metrics}

This appendix defines the full set of metrics used consistently across every phase of this study.

\paragraph{Reciprocal Rank and MRR}
For a query $q$ with ranked list $\pi_q$, reciprocal rank is defined as
\begin{equation}
RR(q) = \frac{1}{r_q}
\label{eq:rr}
\end{equation}
where $r_q$ is the rank position of the first officially relevant candidate in $\pi_q$. Mean Reciprocal Rank (MRR) is the average of $RR(q)$ over all queries:
\begin{equation}
MRR = \frac{1}{|Q|}\sum_{q \in Q} \frac{1}{r_q}
\label{eq:mrr}
\end{equation}
MRR is sensitive only to the position of the first relevant result and is used throughout this study as a quick indicator of whether a system surfaces at least one correct candidate early.

\paragraph{Precision@K and Recall@K.}
For a cutoff $K$, precision and recall at $K$ are computed per query as
\begin{equation}
P@K(q) = \frac{|\pi_q^{(K)} \cap R_q|}{K}, \qquad
R@K(q) = \frac{|\pi_q^{(K)} \cap R_q|}{|R_q|}
\label{eq:precrec}
\end{equation}
where $\pi_q^{(K)}$ denotes the top $K$ candidates in the ranked list and $R_q$ is the officially relevant set for query $q$. These are reported either per query or macro averaged across queries, and are used primarily to characterize candidate pool recall at intermediate pipeline stages (e.g., recall of the top 600 candidate pool).

\paragraph{Mean Average Precision (MAP).}
Average Precision for a single query averages precision at each rank position where a relevant candidate appears:
\begin{equation}
AP(q) = \frac{1}{|R_q|}\sum_{k \,:\, c_k \in R_q} P@k(q)
\label{eq:ap}
\end{equation}
MAP is the mean of $AP(q)$ over all queries. Unlike MRR, MAP accounts for the position of every relevant candidate in the ranking, not just the first.

\paragraph{Normalized Discounted Cumulative Gain (nDCG@K).}
For binary relevance, Discounted Cumulative Gain at rank $K$ is
\begin{equation}
DCG@K(q) = \sum_{i=1}^{K} \frac{rel_i}{\log_2(i+1)}
\label{eq:dcg}
\end{equation}
where $rel_i \in \{0,1\}$ indicates whether the candidate at rank $i$ is officially relevant. nDCG@K normalizes this by the ideal DCG obtainable at that cutoff, $IDCG@K(q)$, the DCG of the best possible ordering:
\begin{equation}
nDCG@K(q) = \frac{DCG@K(q)}{IDCG@K(q)}
\label{eq:ndcg}
\end{equation}
nDCG rewards relevant candidates appearing earlier in the ranking more heavily than those appearing later, and is reported as a mean across queries.

\paragraph{Micro Precision, Micro Recall, and Micro F1.}
These are the primary metrics used for all official evaluation in this study, computed by the frozen evaluator described in the main text. Unlike the per query metrics above, micro averaged metrics aggregate hits and denominators across the entire query set before computing a ratio, rather than averaging per query ratios:
\begin{equation}
P_{\mu}@K = \frac{\sum_{q} |\pi_q^{(K)} \cap R_q|}{\sum_{q} |\pi_q^{(K)}|}, \qquad
R_{\mu}@K = \frac{\sum_{q} |\pi_q^{(K)} \cap R_q|}{\sum_{q} |R_q|}, \qquad
F1_{\mu}@K = \frac{2\,P_{\mu}@K \cdot R_{\mu}@K}{P_{\mu}@K + R_{\mu}@K}
\label{eq:microf1}
\end{equation}
Because these are computed as ratios of pooled sums rather than as an average of per query scores, queries with more officially relevant candidates contribute proportionally more to the final metric. This micro aggregated Precision/Recall/F1 at the frozen cutoff is treated as the single official measure of system quality throughout this study; all other metrics are reported as supporting context.

\paragraph{Hit@K.}
The fraction of query groups with at least one relevant candidate within the top $K$:
\begin{equation}
Hit@K = \frac{1}{|Q|}\sum_{q \in Q} \mathbb{1}\!\left[|\pi_q^{(K)} \cap R_q| \geq 1\right]
\label{eq:hitk}
\end{equation}
Used only in internal group ranking evaluation of intermediate neural rankers, not as an official metric.

\paragraph{Secondary classification level metrics.}
In addition to the ranking metrics above, two intermediate components --- a role combination classifier and an early cross-encoder --- are evaluated at the level of binary pair classification rather than per query ranking, using standard ROC-AUC, binary F1, and Average Precision (precision averaged across recall thresholds for the positive class). These metrics characterize how well a component separates relevant from non relevant pairs in isolation and are not directly comparable to the per query ranking metrics above; they are reported only where they originate, and are never substituted for official Micro F1 in any final comparison.

\paragraph{Macro versus micro aggregation.}
Where both are reported, macro averaged metrics (per query score, then averaged across queries) and micro averaged metrics (pooled sums, then a single ratio) are explicitly distinguished, since the two can diverge substantially --- a query with a large number of officially relevant candidates influences a micro metric far more than a macro metric. All results labeled \emph{official} in this study use the micro aggregated convention, matching the evaluator used for the final reported result.

\end{appendices}


\bibliography{sn-bibliography}

\end{document}